\documentclass{aa}  

\usepackage{textcomp}
\usepackage{graphicx} %
\usepackage{amsmath} %
\usepackage{amssymb} %
\usepackage{bm} %
\usepackage{upgreek} %
\usepackage{IEEEtrantools} %
\usepackage{multirow}
\usepackage[colorlinks=true,allcolors=blue,urlcolor=blue]{hyperref}
\usepackage{txfonts}
\usepackage[normalem]{ulem} 
\usepackage{multirow}

\newcommand{\sect}[1]{\text{Sect.~\ref{#1}}}
\newcommand{\fig}[1]{\text{Fig.~\ref{#1}}}
\newcommand{\tab}[1]{\text{Table~\ref{#1}}}

\newcommand{\multitd}{\textsc{multi3d}}

\newcommand{\blob}{\textsc{balder}}
\newcommand{\mtd}{\textlangle3D\textrangle}

\newcommand{\stagger}{\textsc{stagger}}
\newcommand{\atmo}{\textsc{atmo}}

\newcommand{\kms}{\mathrm{km\,s^{-1}}}

\newcommand{\feh}{\mathrm{\left[Fe/H\right]}}
\newcommand{\abrat}[2]{\mathrm{\left[#1/#2\right]}}

\newcommand{\lgeps}[1]{\log{\epsilon_{\mathrm{#1}}}}

\newcommand{\lgr}{\log{\tau_{\mathrm{R}}}}
\newcommand{\dex}{\mathrm{dex}}

\newcommand{\nm}{\mathrm{nm}}
\begin{document} 

\title{3D non-LTE line formation of neutral carbon in the Sun}
\author{A.~M.~Amarsi\inst{1}
\and
P.~S.~Barklem\inst{2}
\and
R.~Collet\inst{3}
\and
N.~Grevesse\inst{4,5}
\and
M.~Asplund\inst{6}}
\institute{Max Planck Institute f\"ur Astronomy, K\"onigstuhl 17, 
D-69117 Heidelberg, Germany\\
\email{amarsi@mpia.de}
\and
Theoretical Astrophysics, Department of Physics and Astronomy, 
Uppsala University, Box 516, SE-751 20 Uppsala, Sweden
\and
Stellar Astrophysics Centre, Department of Physics and Astronomy, Aarhus
University, Ny Munkegade 120, DK-8000 Aarhus C, Denmark
\and
Centre Spatial de Li\`ege, Universit\'e de Li\'ege, avenue Pr\'e Aily, 4031 Angleur-Li\`ege, Belgium
\and
Space Sciences, Technologies and Astrophysics Research (STAR)
Institute, 
Universit\'e de Li\`ege, All\'ee du 6 ao\^ut, 17, B5C, 4000 Li\`ege, Belgium
\and
Research School of Astronomy and Astrophysics, Australian National
University, Canberra, ACT 2611, Australia}

\abstract{Carbon abundances in late-type stars
are important in a variety of astrophysical contexts.
However \ion{C}{I} lines, one of the main abundance
diagnostics, are sensitive to departures 
from local thermodynamic equilibrium (LTE).
We present a model atom for non-LTE analyses of \ion{C}{I}~lines,
that uses a new, physically-motivated recipe for
the rates of neutral hydrogen impact excitation.
We analyse \ion{C}{I}~lines in the solar spectrum,
employing a three-dimensional (3D) hydrodynamic model solar atmosphere and
3D non-LTE radiative transfer. 
We find negative non-LTE abundance corrections
for \ion{C}{I}~lines
in the solar photosphere, in accordance with previous studies,
reaching up to around $0.1\,\dex$~in the disk-integrated flux.
We also present the first fully consistent 
3D non-LTE solar carbon abundance determination: we infer 
$\lgeps{C}=8.44\pm0.02$,
in good agreement with the current standard value.
Our models reproduce the observed solar centre-to-limb variations
of various \ion{C}{I}~lines, without any adjustments
to the rates of neutral hydrogen impact excitation,
suggesting that the proposed recipe
may be a solution to the long-standing problem
of how to reliably model inelastic collisions with neutral hydrogen
in late-type stellar atmospheres.}


\maketitle

\section{Introduction}
\label{introduction}

As the second most abundant metal at the present epoch, 
carbon is interesting in a variety of astrophysical contexts.
Via the C/O ratio, carbon has an oversized influence
on the structure of the atmospheres \citep[e.g.][]{2012ApJ...758...36M}
and interiors \citep[e.g.][]{2005astro.ph..4214K} of exoplanets. 
Carbon is similarly influential on the structure
of the atmospheres \citep[e.g.][]{2017A&amp;A...598L..10G}
and interiors \citep[e.g.][]{2008PhR...457..217B} of stars,
and helps shape the properties
of early stars \citep[e.g][]{2003Natur.425..812B}.  
Carbon is produced from a number of different sites 
including core collapse supernova \citep[e.g.][]{2002RvMP...74.1015W},
AGB stars \citep[e.g.][]{2014MNRAS.445..347K},
and Wolf-Rayet stars \citep[e.g.][]{2018ApJS..237...13L};
as such the Galactic Chemical Evolution of carbon
is likely to be metallicity dependent
and is still rather uncertain
\citep[e.g.][]{2011MNRAS.414.3231K}.

Thus, accurate carbon abundances measured across
the Galaxy are of great interest to the community.
In late-type stars, \ion{C}{I} lines are among the most commonly used
carbon abundance diagnostics
\citep[e.g.][]{2013PASJ...65...53T,2014A&amp;A...568A..25N,
2016ApJ...833..225Z}, and can be measured down to
$\feh$\footnote{$\abrat{A}{B}=
\left(\log{N_{\mathrm{A}}}/\log{N_{\mathrm{B}}}\right)_{*}-
\left(\log{N_{\mathrm{A}}}/
\log{N_{\mathrm{B}}}\right)_{\odot}$.}$\approx-3.0$~\citep[e.g.
][]{2009A&amp;A...500.1143F}.
However, in late-type stellar atmospheres, \ion{C}{I} is sensitive to 
departures from local thermodynamic equilibrium 
\citep[LTE; e.g.][]{1981ApJS...45..635V,
1990A&A...237..125S,1992PASJ...44..649T,
2005A&amp;A...431..693A,
2005PASJ...57...65T,2006A&amp;A...458..899F,
2015MNRAS.453.1619A}.
For reliable results based on \ion{C}{I} lines, non-LTE effects 
need to be taken into account.

A question to ask of non-LTE studies in general,
is how realistic are the model atoms being used.
For \ion{C}{I}, the main uncertainties have
hitherto been related to the completeness of the model atom
\citep[e.g.][]{1990A&A...237..125S},
and to the treatment of neutral hydrogen impact excitation
\citep[e.g.][]{2006A&amp;A...458..899F,
2010A&amp;A...514A..92C,
2015MNRAS.453.1619A}.
This latter point has lately been given considerable 
attention in the context of other elements.
Lacking collisional cross-sections based on
expensive quantum chemistry calculations of the molecular structure
\citep[e.g.][]{2003PhRvA..68f2703B,
2010PhRvA..81c2706B,2012PhRvA..85c2704B},
works have until recently typically either 
neglected neutral hydrogen impact excitation, or
included them using the Drawin recipe
\citep{1968ZPhy..211..404D,1969ZPhy..225..483D,
1984A&amp;A...130..319S,1993PhST...47..186L}.
This recipe however is not well-founded from a physical
point of view, and is typically in error by orders of magnitude,
with the errors going in different directions
for different transitions 
\citep[e.g.][and references therein]{2016A&amp;ARv..24....9B}.

Recently, new recipes for neutral hydrogen impact excitation
have become available that are
both physically motivated and are relatively inexpensive to compute.
In our study of neutral oxygen
\citep{2018A&A...616A..89A}, we suggested employing
the asymptotic two electron model
of \citet{2016PhRvA..93d2705B}, combined with
the free electron model of \citet{1991JPhB...24L.127K}. 
These two models account for different physical
mechanisms, with the asymptotic model
particularly suitable for low level transitions
while the free electron model 
should perform well for lines from
intermediate- and high-excitation levels.
Since the permitted optical and infrared 
\ion{C}{I} lines are all in this regime, 
we expect that the mechanism represented by
the free electron model is important here.

One way to validate model atoms is to use them in
analyses of the solar spectrum.
Trends in abundances inferred from spectral lines 
of different parameters 
(e.g.~wavelength, line strength, excitation potential)
have traditionally been used to probe errors in the modelling.
More recently, centre-to-limb variations of
spectral lines have been used to probe
different recipes of neutral hydrogen impact excitation 
\citep[][]{2004A&amp;A...423.1109A,
2005ARA&amp;A..43..481A,
2009A&amp;A...508.1403P,2015A&amp;A...583A..57S,
2017MNRAS.468.4311L,2018A&A...616A..89A}.
This is because 
a) spectral lines observed at the limb
form higher up than lines observed at disk-centre, 
b) free electrons are the main thermalising
agents in the deep solar photosphere, and 
c) the relative importance of 
neutral hydrogen to free electrons as thermalising agents
follows the ratio $N_{\mathrm{H}}/N_{\mathrm{e^{-}}}$,
which increases with height.

In order to use the Sun to validate atomic data in this way, 
one requires a reliable method for spectral line synthesis. 
The current state-of-the-art employs three-dimensional
(3D) radiative-hydrodynamic model solar atmospheres,
and 3D non-LTE post-processing radiative transfer 
\citep[e.g.][]{2018A&A...616A..89A}.
To our knowledge, consistent 3D non-LTE calculations for 
\ion{C}{I} in the solar photosphere
that are based on a model atom 
that is sufficiently large for reliable abundance determinations
have not been presented in the literature until now.

Here, we present a new model atom for
\ion{C}{I} line formation (\sect{method}),
that uses a physically-motivated recipe for
the rates of neutral hydrogen impact excitation.
We carry out detailed 3D non-LTE calculations
for \ion{C}{I}~lines in the solar photosphere,
looking at disk-centre abundance trends
as well as centre-to-limb variations
of various clean \ion{C}{I}~lines in order to validate the model atom
(\sect{results}).
We discuss the implications 
on the solar carbon abundance and comment
on the non-LTE effects on \ion{C}{I}~lines (\sect{discussion}).
We close with a short summary (\sect{conclusion}).

\section{Method}
\label{method}

\subsection{Line-list}
\label{methodlinelist}

\begin{table*}
\begin{center}
\caption{Parameters of the full set of lines for which 3D non-LTE effects are discussed. Nominal uncertainties in $\log gf$~were taken from NIST. For the permitted lines, broadening via elastic collisions with hydrogen atoms is represented using $\sigma$, the cross-section at the reference velocity of $10^{4}\,\mathrm{m\,s^{-1}}$, and $\alpha$, the exponent such that the cross-section varies with velocity as $\varv^{-\alpha}$~\citep[e.g.][]{1995MNRAS.276..859A}. }
\label{tab:linelist}
\begin{tabular}{rcl c c c c c c c}
\hline
\multicolumn{3}{c}{Transition} &
\multicolumn{1}{c}{$\lambda_{\text{air}}$} &
\multicolumn{1}{c}{$\chi_{\text{exc.}}$} &
\multicolumn{1}{c}{$\log gf$} &
\multicolumn{1}{c}{$\sigma_{\log gf}$} &
\multicolumn{1}{c}{$\log(\gamma_{\text{rad.}}/s^{-1})$} &
\multicolumn{1}{c}{$\sigma / a_{0}^{2}$} &
\multicolumn{1}{c}{$\alpha$ }\\
\hline
\hline
$\mathrm{2p^{2}\,^{1}D_{2}}$~&---&~$\mathrm{2p^{2}\,^{1}S_{0}}$ &
$  872.7121$ &
$     1.264$ &
$    -8.165$ &
$      0.05$ &
$-     0.070$ &
 &
 \\
\hline
$\mathrm{2p.3s\,^{1}P^{o}_{1}}$~&---&~$\mathrm{2p.4p\,^{1}D_{2}}$ &
$  505.2164$ &
$     7.685$ &
$    -1.303$ &
$      0.05$ &
$+     8.592$ &
$        1041$ &
$    0.2369$ \\
$\mathrm{2p.3s\,^{1}P^{o}_{1}}$~&---&~$\mathrm{2p.4p\,^{1}P_{1}}$ &
$  538.0335$ &
$     7.685$ &
$    -1.616$ &
$      0.05$ &
$+     8.589$ &
$        1041$ &
$    0.2369$ \\
$\mathrm{2p.3p\,^{1}P_{1}}$~&---&~$\mathrm{2p.4d\,^{1}P^{o}_{1}}$ &
$  658.7606$ &
$     8.537$ &
$    -1.003$ &
$      0.05$ &
$+     7.979$ &
$        1944$ &
$    0.3176$ \\
$\mathrm{2p.3p\,^{3}D_{1}}$~&---&~$\mathrm{2p.4d\,^{3}F^{o}_{2}}$ &
$  711.1467$ &
$     8.640$ &
$    -1.085$ &
$      0.05$ &
$+     7.431$ &
$        1849$ &
$    0.3135$ \\
$\mathrm{2p.3p\,^{3}D_{3}}$~&---&~$\mathrm{2p.4d\,^{3}F^{o}_{4}}$ &
$  711.3177$ &
$     8.647$ &
$    -0.773$ &
$      0.05$ &
$+     7.431$ &
$        1864$ &
$    0.3141$ \\
$\mathrm{2p.3s\,^{1}P^{o}_{1}}$~&---&~$\mathrm{2p.3p\,^{1}S_{0}}$ &
$  833.5141$ &
$     7.685$ &
$    -0.437$ &
$      0.03$ &
$+     8.630$ &
$         690$ &
$    0.2257$ \\
$\mathrm{2p.3s\,^{3}P^{o}_{1}}$~&---&~$\mathrm{2p.3p\,^{3}P_{2}}$ &
$  906.1427$ &
$     7.483$ &
$    -0.347$ &
$      0.05$ &
$+     8.595$ &
$         581$ &
$    0.2343$ \\
$\mathrm{2p.3s\,^{3}P^{o}_{0}}$~&---&~$\mathrm{2p.3p\,^{3}P_{1}}$ &
$  906.2481$ &
$     7.480$ &
$    -0.455$ &
$      0.05$ &
$+     8.595$ &
$         580$ &
$    0.2343$ \\
$\mathrm{2p.3s\,^{3}P^{o}_{1}}$~&---&~$\mathrm{2p.3p\,^{3}P_{1}}$ &
$  907.8278$ &
$     7.483$ &
$    -0.581$ &
$      0.05$ &
$+     8.595$ &
$         580$ &
$    0.2343$ \\
$\mathrm{2p.3s\,^{3}P^{o}_{1}}$~&---&~$\mathrm{2p.3p\,^{3}P_{0}}$ &
$  908.8503$ &
$     7.483$ &
$    -0.430$ &
$      0.05$ &
$+     8.595$ &
$         580$ &
$    0.2344$ \\
$\mathrm{2p.3s\,^{3}P^{o}_{2}}$~&---&~$\mathrm{2p.3p\,^{3}P_{2}}$ &
$  909.4824$ &
$     7.488$ &
$+     0.151$ &
$      0.05$ &
$+     8.595$ &
$         581$ &
$    0.2343$ \\
$\mathrm{2p.3s\,^{3}P^{o}_{2}}$~&---&~$\mathrm{2p.3p\,^{3}P_{1}}$ &
$  911.1800$ &
$     7.488$ &
$    -0.297$ &
$      0.05$ &
$+     8.595$ &
$         580$ &
$    0.2344$ \\
$\mathrm{2p.3s\,^{1}P^{o}_{1}}$~&---&~$\mathrm{2p.3p\,^{1}D_{2}}$ &
$  940.5721$ &
$     7.685$ &
$+     0.286$ &
$      0.03$ &
$+     8.620$ &
$         633$ &
$    0.2271$ \\
$\mathrm{2p.3s\,^{3}P^{o}_{0}}$~&---&~$\mathrm{2p.3p\,^{3}S_{1}}$ &
$  960.3027$ &
$     7.480$ &
$    -0.896$ &
$      0.05$ &
$+     8.586$ &
$         562$ &
$    0.2363$ \\
$\mathrm{2p.3s\,^{3}P^{o}_{1}}$~&---&~$\mathrm{2p.3p\,^{3}S_{1}}$ &
$  962.0767$ &
$     7.483$ &
$    -0.445$ &
$      0.05$ &
$+     8.586$ &
$         562$ &
$    0.2363$ \\
$\mathrm{2p.3s\,^{3}P^{o}_{2}}$~&---&~$\mathrm{2p.3p\,^{3}S_{1}}$ &
$  965.8423$ &
$     7.488$ &
$    -0.280$ &
$      0.05$ &
$+     8.586$ &
$         562$ &
$    0.2364$ \\
$\mathrm{2p.3p\,^{1}P_{1}}$~&---&~$\mathrm{2p.3d\,^{1}P^{o}_{1}}$ &
$  1012.387$ &
$     8.537$ &
$    -0.031$ &
$      0.09$ &
$+     8.264$ &
$         822$ &
$    0.2740$ \\
$\mathrm{2p.3s\,^{3}P^{o}_{0}}$~&---&~$\mathrm{2p.3p\,^{3}D_{1}}$ &
$  1068.534$ &
$     7.480$ &
$    -0.272$ &
$      0.05$ &
$+     8.578$ &
$         534$ &
$    0.2377$ \\
$\mathrm{2p.3s\,^{3}P^{o}_{1}}$~&---&~$\mathrm{2p.3p\,^{3}D_{1}}$ &
$  1070.731$ &
$     7.483$ &
$    -0.411$ &
$      0.05$ &
$+     8.578$ &
$         534$ &
$    0.2377$ \\
$\mathrm{2p.3s\,^{3}P^{o}_{2}}$~&---&~$\mathrm{2p.3p\,^{3}D_{2}}$ &
$  1072.952$ &
$     7.488$ &
$    -0.420$ &
$      0.05$ &
$+     8.578$ &
$         535$ &
$    0.2379$ \\
$\mathrm{2p.3s\,^{3}P^{o}_{2}}$~&---&~$\mathrm{2p.3p\,^{3}D_{1}}$ &
$  1075.397$ &
$     7.488$ &
$    -1.606$ &
$      0.05$ &
$+     8.578$ &
$         534$ &
$    0.2378$ \\
$\mathrm{2p.3p\,^{3}D_{2}}$~&---&~$\mathrm{2p.3d\,^{3}F^{o}_{2}}$ &
$  1177.753$ &
$     8.643$ &
$    -0.520$ &
$      0.05$ &
$+     7.760$ &
$         748$ &
$    0.2704$ \\
$\mathrm{2p.3p\,^{3}D_{2}}$~&---&~$\mathrm{2p.4s\,^{3}P^{o}_{2}}$ &
$  1184.870$ &
$     8.643$ &
$    -0.697$ &
$      0.05$ &
$+     8.143$ &
$        1387$ &
$    0.2221$ \\
$\mathrm{2p.3p\,^{3}D_{1}}$~&---&~$\mathrm{2p.4s\,^{3}P^{o}_{1}}$ &
$  1186.300$ &
$     8.640$ &
$    -0.710$ &
$      0.05$ &
$+     8.143$ &
$        1380$ &
$    0.2217$ \\
$\mathrm{2p.3p\,^{3}D_{2}}$~&---&~$\mathrm{2p.4s\,^{3}P^{o}_{1}}$ &
$  1189.289$ &
$     8.643$ &
$    -0.277$ &
$      0.05$ &
$+     8.143$ &
$        1381$ &
$    0.2216$ \\
$\mathrm{2p.3p\,^{3}D_{3}}$~&---&~$\mathrm{2p.4s\,^{3}P^{o}_{2}}$ &
$  1189.575$ &
$     8.647$ &
$    -0.008$ &
$      0.05$ &
$+     8.143$ &
$        1388$ &
$    0.2220$ \\
$\mathrm{2p.3p\,^{3}P_{0}}$~&---&~$\mathrm{2p.3d\,^{3}P^{o}_{1}}$ &
$  1254.949$ &
$     8.847$ &
$    -0.565$ &
$      0.05$ &
$+     8.375$ &
$         866$ &
$    0.2935$ \\
$\mathrm{2p.3p\,^{3}P_{1}}$~&---&~$\mathrm{2p.3d\,^{3}P^{o}_{0}}$ &
$  1256.212$ &
$     8.848$ &
$    -0.522$ &
$      0.05$ &
$+     8.375$ &
$         867$ &
$    0.2936$ \\
$\mathrm{2p.3p\,^{3}P_{1}}$~&---&~$\mathrm{2p.3d\,^{3}P^{o}_{1}}$ &
$  1256.904$ &
$     8.848$ &
$    -0.598$ &
$      0.05$ &
$+     8.375$ &
$         866$ &
$    0.2935$ \\
$\mathrm{2p.3p\,^{3}P_{1}}$~&---&~$\mathrm{2p.3d\,^{3}P^{o}_{2}}$ &
$  1258.158$ &
$     8.848$ &
$    -0.536$ &
$      0.05$ &
$+     8.375$ &
$         865$ &
$    0.2932$ \\
$\mathrm{2p.3p\,^{1}D_{2}}$~&---&~$\mathrm{2p.4s\,^{1}P^{o}_{1}}$ &
$  1744.856$ &
$     9.003$ &
$+     0.012$ &
$      0.03$ &
$+     8.052$ &
$        1491$ &
$    0.2727$ \\
$\mathrm{2p.3p\,^{1}S_{0}}$~&---&~$\mathrm{2p.3d\,^{1}P^{o}_{1}}$ &
$  2102.316$ &
$     9.172$ &
$    -0.398$ &
$      0.05$ &
$+     8.339$ &
$         788$ &
$    0.2804$ \\
$\mathrm{2p.3p\,^{1}S_{0}}$~&---&~$\mathrm{2p.4s\,^{1}P^{o}_{1}}$ &
$  2290.654$ &
$     9.172$ &
$    -0.217$ &
$      0.05$ &
$+     8.086$ &
$        1382$ &
$    0.3353$ \\
$\mathrm{2p.3d\,^{1}F^{o}_{3}}$~&---&~$\mathrm{2p.4p\,^{1}D_{2}}$ &
$  3085.412$ &
$     9.736$ &
$+     0.097$ &
$      0.05$ &
$+     8.393$ &
$         920$ &
$    0.2793$ \\
$\mathrm{2p.4p\,^{1}P_{1}}$~&---&~$\mathrm{2p.4d\,^{1}D^{o}_{2}}$ &
$  3406.491$ &
$     9.989$ &
$+     0.443$ &
$      0.05$ &
$+     7.319$ &
$        1653$ &
$    0.2653$ \\
\hline
\hline
\end{tabular}
\end{center}
\end{table*}

\begin{figure}
    \begin{center}
        \includegraphics[scale=0.31]{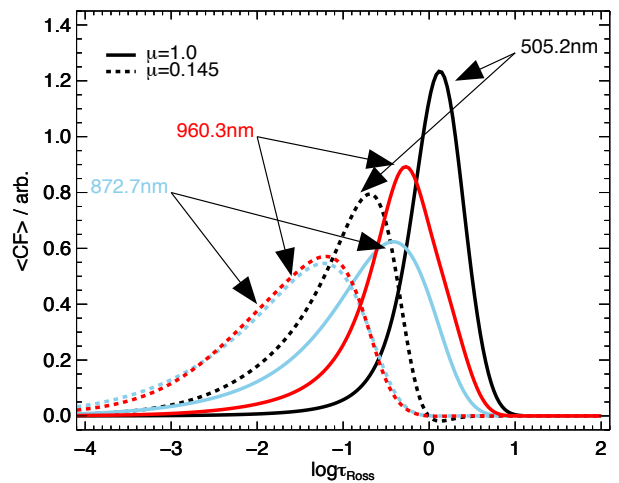}
        \caption{Contribution functions to the intensity depression for the 
        \ion{C}{I}~$505.2\,\nm$~and $960.3\,\nm$~lines,
        and the [\ion{C}{I}]~$872.7\,\nm$~line,
        integrated over wavelength, at two different viewing angles.
        The contribution functions have been normalised
        such that under each, the area is equal to unity.}
        \label{fig:cf}
    \end{center}
\end{figure}

In \tab{tab:linelist}~we list the 35~\ion{C}{I} lines and single
[\ion{C}{I}] line for which 3D non-LTE line formation
for the Sun were calculated and discussed in detailed (\sect{resultseffects}).
This list was initially constructed by combining the set of lines given in
Table~2 of \citet[][]{2005A&amp;A...431..693A},
with the set of lines assigned a rank of $1$~in
Table~1 of \citet[][]{2010A&amp;A...514A..92C}.
The \ion{C}{I}~$801.9\,\nm$~line was
removed from this list on account of being
very weak. The \ion{C}{I}~$1750.6\,\nm$~line
was also removed, for having a very high excitation potential:
the upper level of this transition was collapsed into a super level
in the model atom, (\sect{methodatomreduced}), making the 
non-LTE effects predicted for this line less reliable. 
The \ion{C}{I} $906.2\,\nm$, $908.9\,\nm$,
$909.5\,\nm$, $940.6\,\nm$,
and $962.1\,\nm$~were added to this list because
they are useful diagnostics in the metal-poor regime
\citep[][]{2009A&amp;A...500.1143F,
2019A&A...622L...4A}.
We adopted different line parameters to those listed in
\citet[][]{2005A&amp;A...431..693A}
and \citet[][]{2010A&amp;A...514A..92C} --
in particular, using the oscillator strengths listed in
the NIST Atomic Spectra Database (\sect{methodatom});
the differences in the oscillator strengths
are at most around $0.02\,\dex$.

For studying disk-centre abundance trends
(\sect{resultstrends}),
it is necessary to avoid lines that may be
affected by solar and telluric blends
in order to avoid biasing the results.
Therefore, a subset of the lines in \tab{tab:linelist} was used.
This subset is composed of $14$~of the $16$~\ion{C}{I}~lines 
as well as the single [\ion{C}{I}]~line presented in
Table~2 of \citet[][]{2005A&amp;A...431..693A};
we opted to remove the \ion{C}{I} $960.3\,\nm$~line
because it is too strong to be a reliable abundance indicator,
and the \ion{C}{I} $2290.6\,\nm$~line
because of difficulties in measuring its equivalent width.
The adopted equivalent widths are those 
used in \citet{2009ARA&amp;A..47..481A}
(updated from \citealt{2005A&amp;A...431..693A}), except for two lines:
the \ion{C}{I}~$711.1\,\nm$~and $711.3\,\nm$~lines.
These two lines are blended with faint CN lines;
the new equivalent widths (\sect{resultstrends})
have taken these blends into account 
($0.08\,\mathrm{pm}$~and $0.20\,\mathrm{pm}$~respectively;
J.~Melendez priv.~comm.).

For studying centre-to-limb variations
(\sect{resultsclv}),
another subset of the lines in \tab{tab:linelist} was used,
due to the availability of high quality observational data.
This subset is composed of 
the \ion{C}{I}~$505.2\,\nm$,
$538.0\,\nm$, $658.8\,\nm$,
$711.1\,\nm$, $711.3\,\nm$,
$833.5\,\nm$, $960.3\,\nm$,
and $965.8\,\nm$~lines.
Most of the infrared lines from \tab{tab:linelist} were not included
because the analysis was based on
the ``SS3'' atlas of \citet{2015A&amp;A...573A..74S}
that only extends to $1000\,\nm$.
The other \ion{C}{I}~lines missing from this selection
are, toward the limb, too contaminated by blends 
or are too difficult
to continuum-normalise because of surrounding blends,
to be useful here.

To aid intuition, in \fig{fig:cf}~we show the contribution functions 
of the intensity depression of the 
\ion{C}{I}~$505.2\,\nm$~($\chi_{\mathrm{exc.}}=7.7\,\mathrm{eV}$)
and $960.3\,\nm$~($\chi_{\mathrm{exc.}}=7.5\,\mathrm{eV}$)
lines,
and the [\ion{C}{I}]~$872.7\,\nm$~
($\chi_{\mathrm{exc.}}=1.3\,\mathrm{eV}$) line,
at the two different
$\mu$~values ($\mu=1.0$~and $\mu=0.145$)
considered in the centre-to-limb variation analysis
(\sect{resultsclv}). These were computed in the
same way as in Sect.~2.1.3~of \citet[][]{2018A&A...616A..89A};
namely, by computing the full 3D contribution function
\citep[][integrand of Eq.~12]{2015MNRAS.452.1612A},
after averaging over surfaces of equal optical depth and
over time, and integrating over wavelength.


\subsection{Radiative transfer code}
\label{methodcode}

The 3D non-LTE radiative transfer code 
\blob~\citep{2018A&A...615A.139A} was used.
The code originated from
\multitd~\citep{1999ASSL..240..379B,2009ASPC..415...87L},
but has been developed and modified,
utilising a new opacity package and other improvements 
for stellar abundance analyses.
In this work, the underlying algorithm
that solves the statistical equilibrum was replaced,
from that presented in Sect.~2.3 of \citet{1992A&amp;A...262..209R},
to that presented in Sect.~2.4 of that same paper.
This was motivated by numerical issues in certain cases,
namely of overlapping radiative transitions combined with
relatively inefficient collisional coupling.
The calculations on 1D model atmospheres
were performed in the same way as the
calculations on 3D model atmospheres,
the only difference being
the inclusion of microturbulent broadening
in the 1D case \citep[e.g.][Chapter 17]{2008oasp.book.....G}.


\subsection{Model atmospheres}
\label{methodatmosphere}

The 3D non-LTE radiative transfer
calculations were performed in post-processing
across eight snapshots, equally spaced in $21~\text{hours}$~of
solar time, of
the same 3D radiative-hydrodynamic model solar atmosphere
that was used in \citet{2018A&A...616A..89A},
calculated using the 
\stagger~code~\citep[][]{Nordlund:1995,1998ApJ...499..914S,
2011JPhCS.328a2003C}.
The mean effective temperature of the whole $21~\text{hour}$~sequence
$5773\,\mathrm{K}$, with a standard deviation of $16\,\mathrm{K}$;
the mean effective temperature of the eight snapshots
is also $5773\,\mathrm{K}$.
For comparison the reference solar effective temperature
is $5772\,\mathrm{K}$~\citep{2016AJ....152...41P}.

The calculations were performed for various carbon abundances 
in steps of $0.2\,\dex$, around a
central value of $\lgeps{C}=8.43$~\citep{2009ARA&amp;A..47..481A}.
Repeating the abundance analysis using steps of $0.4\,\dex$~instead,
resulted in inferred abundances that changed by at most
$0.01\,\dex$~(for the [\ion{C}{I}]~$872.7\,\nm$~line).

For testing purposes, post-processing radiative transfer 
calculations were performed on the 
horizontally- and temporally-averaged 3D model solar atmosphere
(hereafter \mtd).  The averaging was performed over
the full $21\,\text{hour}$~sequence.
This is the same \mtd~model solar atmosphere 
that was used in \citet{2018A&A...616A..89A}.
For interpreting the 3D effects, radiative transfer 
calculations were also performed on a 1D hydrostatic model 
solar atmosphere, which was calculated
using the \atmo~code \citep[][Appendix A]{2013A&amp;A...557A..26M}.
The \atmo~model was constructed
assuming a microturbulence of $\xi=1.0\,\kms$,
and using the same equation of state and opacity binning scheme
as used for the \stagger~model solar 
atmosphere, permitting a fair comparison.

For both the 1D and the \mtd~model solar atmospheres, 
the post-processing radiative
transfer calculations were performed for the
same carbon abundances as were used on the 3D model solar
atmosphere (i.e.~in steps of $0.2\,\dex$, around a
central value of $\lgeps{C}=8.43$). 
These calculations were performed for a single,
depth-independent microturbulence of $\xi=1.0\,\kms$.

\subsection{Model atom}
\label{methodatom}

\subsubsection{Overview}
\label{methodatomoverview}

\begin{figure*}
    \begin{center}
        \includegraphics[scale=0.31]{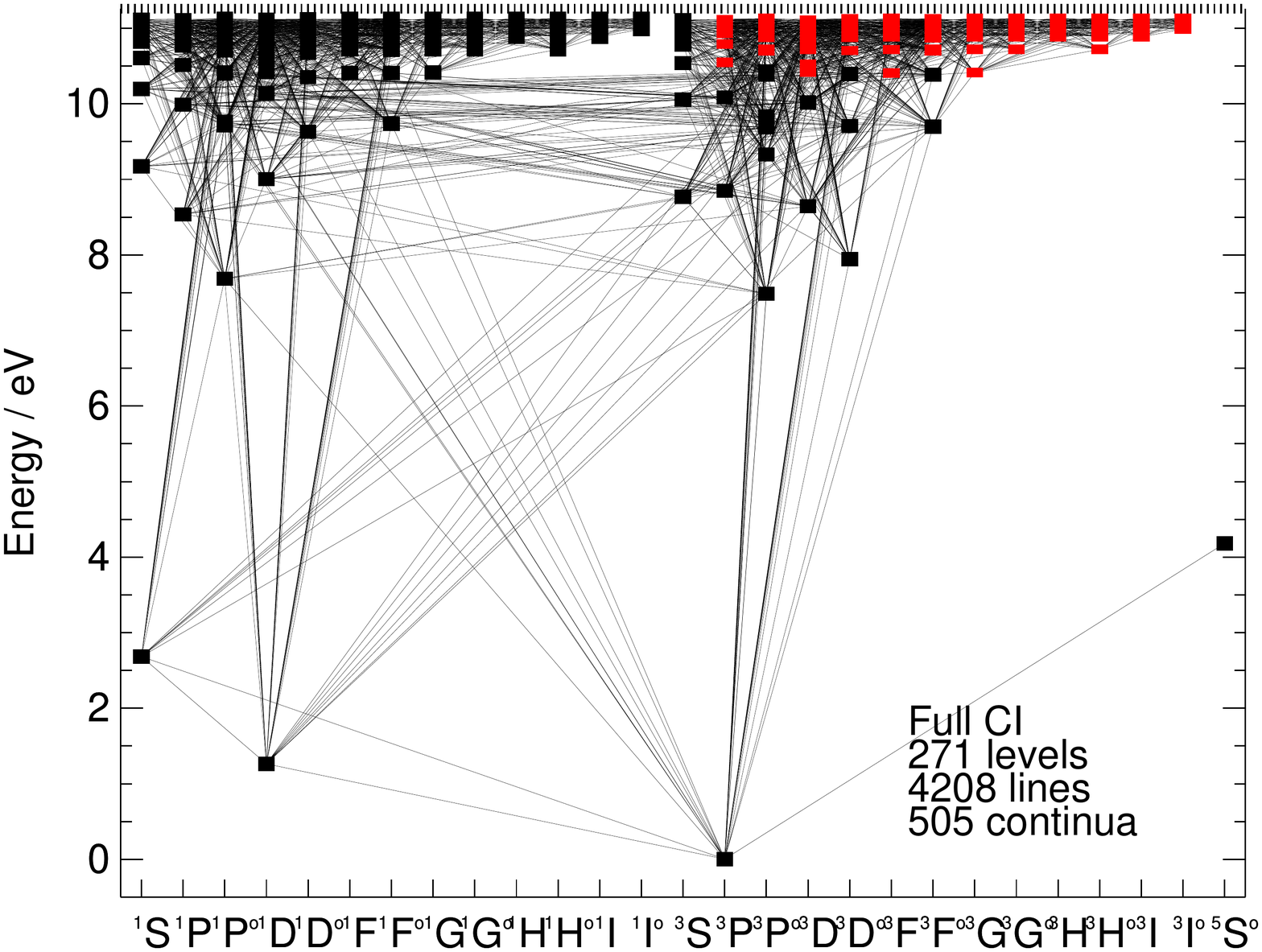}
        \includegraphics[scale=0.31]{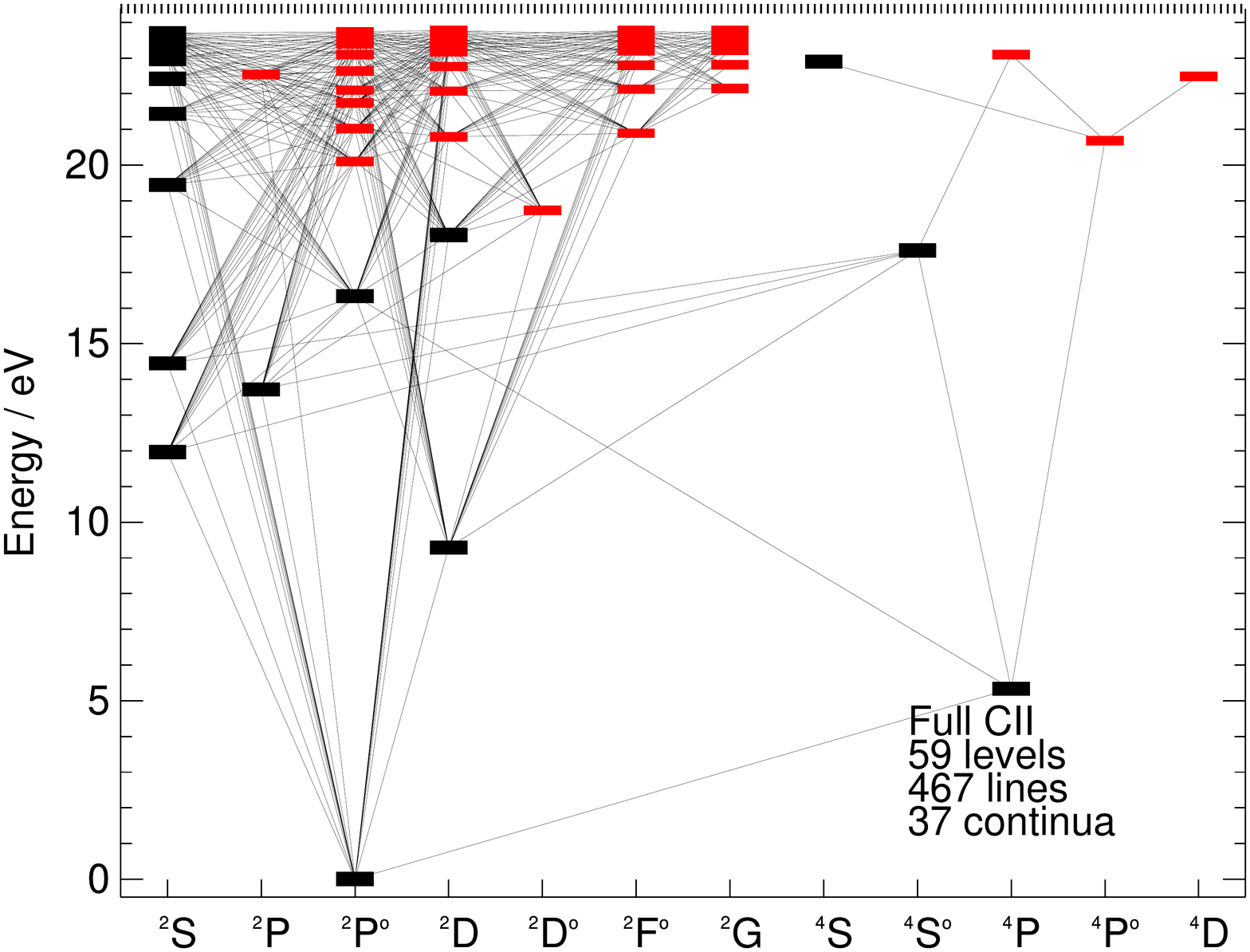}
        \caption{Grotrian diagrams for \ion{C}{I} and \ion{C}{II}
        in the comprehensive model atom.  Levels 
        that do not resolve fine structure are shown as red lines.}
        \label{fig:grotrian_comp}
    \end{center}
\end{figure*}

\begin{figure*}
    \begin{center}
        \includegraphics[scale=0.62]{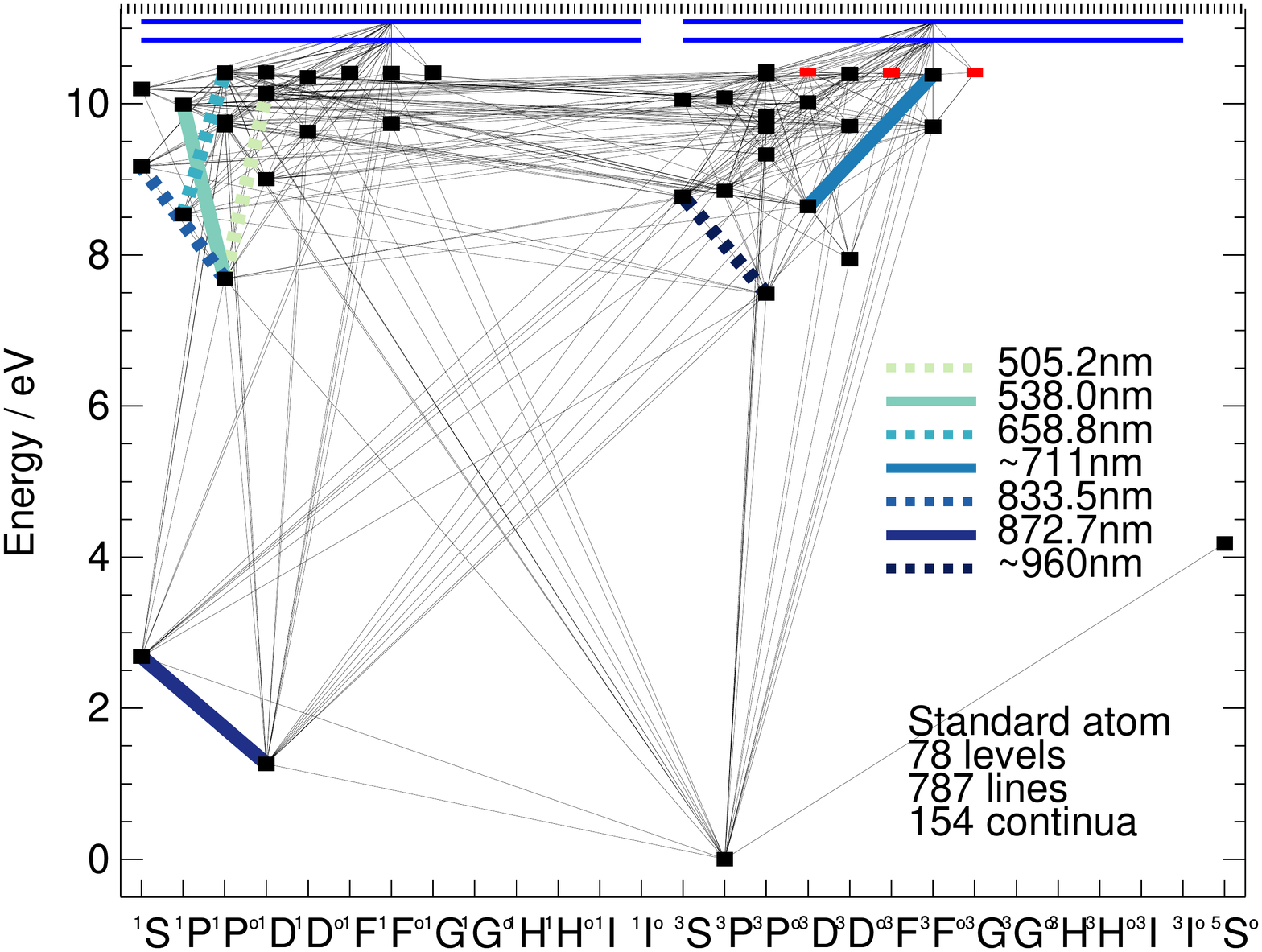}
        \caption{Grotrian diagram for \ion{C}{I}
        in the standard model atom, after
        reducing the comprehensive model by
        removing \ion{C}{III} and
        all excited levels of \ion{C}{II},
        and by constructing super levels.
        Levels 
        that do not resolve fine structure are shown as red lines,
        and super levels are shown as long horizontal lines.
        Some of the \ion{C}{I} lines explicitly considered in this work 
        have been highlighted;
        ``$\sim711\,\nm$''~refers to the
        \ion{C}{I}~$711.1\,\nm$~and 
        $711.3\,\nm$~lines,
        and ``$\sim960\,\nm$''~refers to the 
        \ion{C}{I}~$960.3\,\nm$~and 
        $965.8\,\nm$~lines.}
        \label{fig:grotrian_red}
    \end{center}
\end{figure*}

In this section we focus on the 
construction and reduction of the model atom for \ion{C}{I}.
The procedure is similar to that presented in
\citet{2018A&A...616A..89A} for \ion{O}{I}. 
The model atom was constructed in two steps.
In the first step, a comprehensive model atom was constructed,
which we illustrate in \fig{fig:grotrian_comp}.
In the second step, the comprehensive model was
reduced in complexity (\sect{methodatomreduced}), 
to generate what we refer to as the standard model atom,
which we illustrate in \fig{fig:grotrian_red}.
Reducing the size of the model atom is necessary
to make the calculations presented here feasible;
nevertheless the reduced model atom still encapsulates
the relevant physics.
In addition to the standard model atom,
further models were used to test the sensitivity
of the results to different ingredients and assumptions
(\sect{methodatomsimple}).

In all cases, 
after the non-LTE iterations were completed on a
reduced model atom, the final emergent intensities
were calculated by applying the departure coefficients
to the comprehensive model atom, taking care
to preserve the total population number.
In this way the final emergent intensities
were based on accurate partition functions and
the correct fine structure splitting,
for higher accuracy.

\subsubsection{Energy levels}
\label{methodatomenergies}

We illustrate the structure of the comprehensive model atom in
\fig{fig:grotrian_comp}.  
Experimental fine structure energies 
were taken from the compilation of \citet{gallagher1993tables}
via the NIST Atomic Spectra Database \citep{NIST_ASD},
and, to make the model more complete,
theoretical energies calculated under the assumption
of pure LS coupling, from \citet{1989JPhB...22.3377L} for \ion{C}{I} and
\citet{1985JPhB...18.2587B} for \ion{C}{II} 
via the Opacity Project online database
\citep[TOPbase;][]{1993A&amp;A...275L...5C}, were also used. 
Care was taken not to duplicate levels in the model,
by using the NIST set only for lower lying levels
(up to $10.43\,\mathrm{eV}$~above the ground
state for \ion{C}{I}, and $18.05\,\mathrm{eV}$~above the ground
state for \ion{C}{II}) and the TOPbase set for the remainder.  
The TOPbase energies were not extracted directly: instead,
they were calculated from the stipulated
effective principal quantum numbers, 
combined with the experimental ionisation limits from NIST.
The \ion{C}{III}~ground state was also 
included in the comprehensive model atom.

\subsubsection{Radiative transitions}
\label{methodatomradiation}

Fine structure oscillator strengths
were taken from \citet{1993A&AS...99..179H},
\citet{2001CaJPh..79..955T}, and
\citet{2006JPhB...39.2159F} for \ion{C}{I}, and 
\citet{2000JPhB...33.2419T} and
\citet{1981A&A....96...91N} for \ion{C}{II}, via NIST. 
For completeness this set was merged with 
theoretical oscillator strengths calculated under the assumption
of pure LS coupling, from \citet{1989JPhB...22.3377L}
for \ion{C}{I} and \citet{1987JPhB...20.6399Y}
for \ion{C}{II}, via TOPbase.
Photoionisation cross-sections were also taken from TOPbase,
and smoothed over using a Gaussian filter
\citep[e.g.][]{2008ApJ...682.1376B}.
For all bound-bound transitions, natural broadening coefficients
were calculated from the lifetimes stipulated in TOPbase.
Where possible, collision broadening due to neutral hydrogen
was calculated 
after the theory of Anstee, Barklem, and O'Mara (ABO),
by interpolating the tables presented in \citet{1995MNRAS.276..859A},
\citet{1997MNRAS.290..102B}, and \citet{1998MNRAS.296.1057B}.

Where possible, the TOPbase radiative 
transitions that connect fine structure level(s)
were split, using the relative strengths tabulated in
Sect.~27 of \citet[][]{1973asqu.book.....A}.
These tables are valid in the LS coupling regime,
which is expected to be a good approximation for \ion{C}{I}.

\subsubsection{Collisional transitions}
\label{methodatomcollision}

For electron impact excitation of \ion{C}{I},
the rate coefficients presented by \citet{2013PhRvA..87a2704W}
calculated under the B-spline R-matrix formalism
\citep[BSR e.g.][]{2006CoPhC.174..273Z}
were adopted where possible.
The data set is complete up to 
$\mathrm{2s^{2}\,2p\,3d\,^{3}P^{o}}$
($9.83\,\mathrm{eV}$~above the ground state).
The BSR data of \citet{2013PhRvA..87a2704W} were also used
for electron impact ionisation of \ion{C}{I}.
Comparisons of BSR with independent methods for other species
signal the reliability of this approach
\citep[e.g][]{2017A&A...606A..11B}.
Where the BSR data set is incomplete, the semi-empirical recipe
of \citet{1962ApJ...136..906V} was used for excitation
(imposing a minimum oscillator strength of $0.05$),
while the empirical formula presented in \citet[][]{1973asqu.book.....A}
was used for ionisation.

For electron impact excitation of \ion{C}{II},
included in the comprehensive model atom,
the cross-sections presented by
\cite{2005A&amp;A...432..731W}, calculated
under the close-coupling R-matrix formalism
\citep[e.g.][]{1976AdAMP..11..143B},
were adopted where possible.
As for \ion{C}{I}, this data set was completed using
the semi-empirical recipe
of \citet{1962ApJ...136..906V},
while the empirical formula presented in \citet[][]{1973asqu.book.....A}
was used for electron impact ionisation of \ion{C}{II}.

As we mentioned in \sect{methodatomoverview},
we adopt a new approach for 
neutral hydrogen impact excitation of \ion{C}{I}.
Following the approach of \citet{2018A&A...616A..89A} for \ion{O}{I},
the rate coefficients predicted by the asymptotic two-electron model,
based on linear combinations of atomic orbitals (``LCAO''),
of \citet{2016PhRvA..93d2705B} were added to
the rate coefficients predicted by the free electron model (``Free''),
of \citet{1991JPhB...24L.127K}.
We refer to this combined approach as ``LCAO+Free''.
The \citet{2017ascl.soft01005B} code was used to determine
the rate coefficients in the latter case; these were
redistributed among spin states following 
Eq.~8 and Eq.~9 of \citet[][]{2016A&amp;ARv..24....9B}.
The asymptotic two electron model was also used to calculate 
charge transfer between \ion{C}{I} and neutral hydrogen
($\mathrm{C+H\leftrightarrow C^{+}+H^{-}}$).
Neutral hydrogen impact ionisation 
($\mathrm{C+H\leftrightarrow C^{+}+e^{-}+H}$)
was also included, through Eq.~8 of \citet{1985JPhB...18L.167K},
but was found to be of much lower importance.

Neutral hydrogen impact excitation of \ion{C}{II} was 
neglected in the model.
Neglecting these processes may in principle lead to 
overestimating the departures from LTE,
and increase the influence of the statistical equilibrium of \ion{C}{II},
on the \ion{C}{I} line strengths.
Ultimately however, we found that the overall influence of \ion{C}{II} 
on \ion{C}{I} lines is anyway small,
(\sect{methodatomreduced}),
so the exact treatment of these processes is unimportant here.

\subsubsection{Standard model atom}
\label{methodatomreduced}

The comprehensive model atom in \fig{fig:grotrian_comp}
is too large to use in full 3D non-LTE radiative transfer 
calculations.  It was therefore reduced, following a similar procedure
to that described in \citet{2018A&A...616A..89A}
for our \ion{O}{I} model atom.  The reduction was performed in several
steps as we outline below.

First, all of the excited \ion{C}{II}~levels 
as well as the \ion{C}{III} level were removed.
Tests revealed that these levels have only a very small impact
on the \ion{C}{I} line strengths:
for the lines listed in \tab{tab:linelist},
the differences in the equivalent widths in the emergent flux are 
negligible in the \mtd~model solar atmosphere
(much less than $0.0001\,\dex$).
Previous studies have included several of the 
lower-lying \ion{C}{II} levels
(e.g.~\citealt{2006A&amp;A...458..899F}
and \citealt{2015MNRAS.453.1619A} include the nine lowest
\ion{C}{II} levels, up to around $18\,\mathrm{eV}$~above the 
\ion{C}{II} ground state), which is more accurate but 
apparently not essential.
The relative insensitivity of \ion{C}{II} on the
statistical equilibrium of \ion{C}{I} can be understood
from the high ionisation energy 
of \ion{C}{I} ($11.26\,\mathrm{eV}$), which
ensures that \ion{C}{I} is the majority species in the solar 
photosphere; also, 
the first excited state of \ion{C}{II} has a large separation
from the ground state ($5.33\,\mathrm{eV}$),
so that only the ground state of \ion{C}{II} needs to be considered here.

Second, high-excitation \ion{C}{I} levels
($10.5\,\mathrm{eV}$~above the ground state)
that are closely separated in energy 
(within $0.5\,\mathrm{eV}$~of each other) and that are 
within the same spin system, were collapsed into super levels.
This is valid under the assumption that the composite levels 
are collisionally-coupled and therefore
have identical departure coefficients.
Within a super level, energies were weighted
using their Boltzmann factors, adopting a temperature
of $5000\,\mathrm{K}$~that is characteristic of the line formation
region. Transitions connected to 
super levels were collapsed into super transitions,
in the way described in \citet{2017MNRAS.464..264A}.
In the \mtd~model solar atmosphere
this reduction has an impact of around $0.003\,\dex$~on
the equivalent width in the flux in the worst case,
the \ion{C}{I}~$3406.5\,\nm$~line of high excitation potential;
for the other lines the impact is less than 
$0.001\,\dex$.

The resulting, reduced model atom 
is the standard atom used in this work.
Abundance results and centre-to-limb variations
are based on this model atom, unless otherwise indicated.
We illustrate the model atom in \fig{fig:grotrian_red}.

\subsubsection{Test atoms}
\label{methodatomsimple}

For testing purposes, a further reduction
was made to the standard model atom described in
\sect{methodatomreduced}: namely,
fine structure was collapsed
following the method described in, for example, \citet{2017MNRAS.464..264A}.
We refer to this as the ``No FS'' model.
In the \mtd~model solar atmosphere
the equivalent widths in the fluxes of the lines
were affected by up to $0.008\,\dex$.
The error is lowest in the vertical intensity
and grows for more inclined rays as the line formation moves outwards.
For example, for the \ion{C}{I}~$960.3\,\nm$~line that forms
over a region extending high up in the atmosphere (\fig{fig:cf}),
the error for $\mu=1.0$~is $0.006\,\dex$, while the error at 
$\mu=0.145$~is almost $0.03\,\dex$.
In general,
for abundance determinations from disk-integrated observations of
late-type stars,
errors of this magnitude are not significant;
\citep[for this reason, the ``No FS'' atom was used in our recent study
of halo stars;][]{2019A&A...622L...4A}.
However, for studying solar centre-to-limb variations,
it is important to retain fine structure in the model atom.

To quantify the influence of neutral hydrogen
impact excitation, a further model atom was constructed,
wherein all neutral hydrogen impact excitation rate coefficients
were set to zero. 
Given that fine structure has only a small impact on the 
statistical equilibrium, to reduce the computational cost
fine structure was also collapsed in this test atom.
We refer to this as the ``No FS / C+H exc.'' model.

\section{Results}
\label{results}

\subsection{Departure coefficients}
\label{resultsdeparture}

\begin{figure}
    \begin{center}
        \includegraphics[scale=0.31]{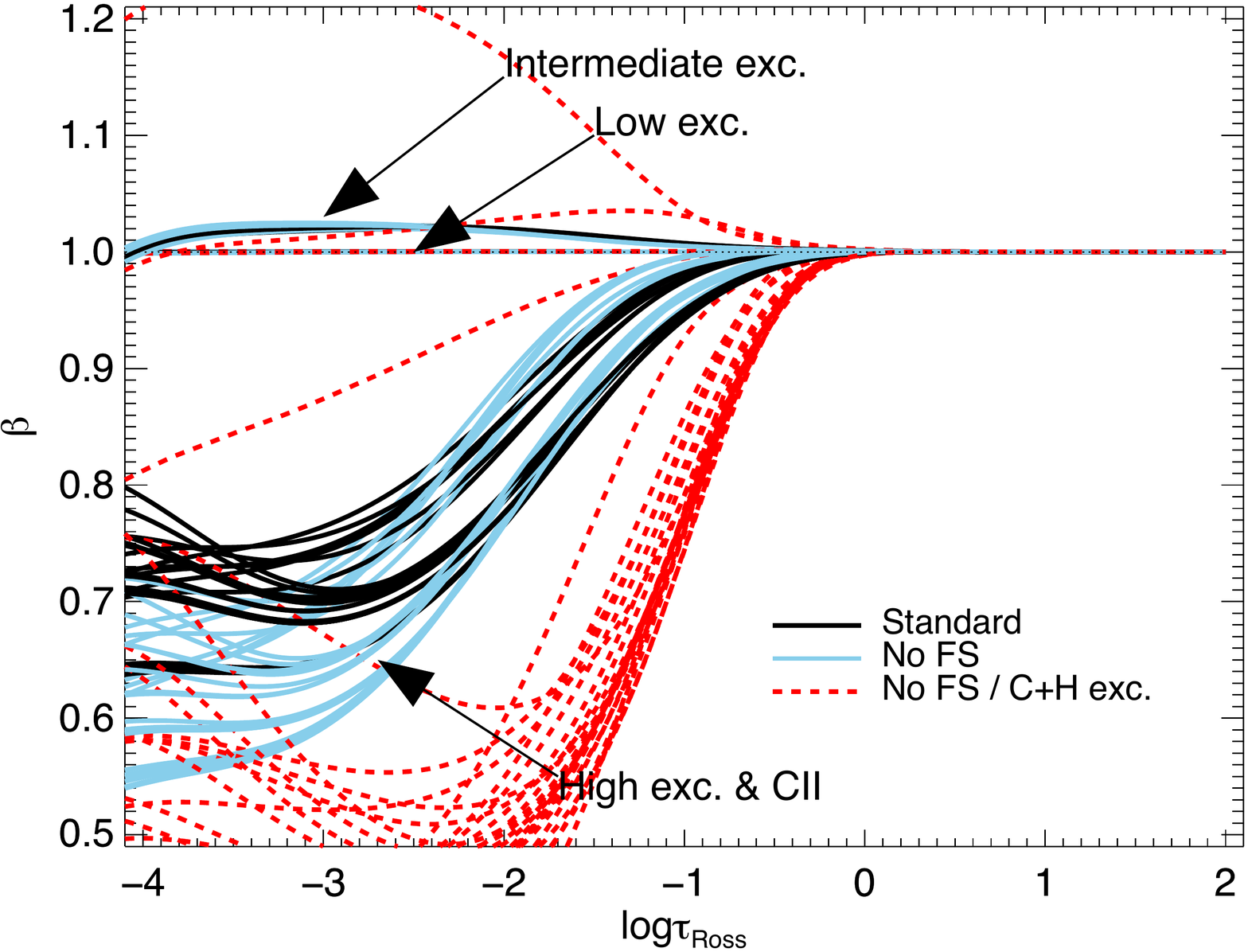}
        \caption{Departure coefficients across the \mtd~model atmosphere,
        for the lowest 42 terms of \ion{C}{1} (up to
        $\mathrm{2s^{2}\,2p\,4d\,^{3}P^{o}}$) as well as
        the ground state of \ion{C}{II}.
        A solar carbon abundance of $\lgeps{C}=8.43$~was assumed.
        The low-excitation levels indicated in the plot span
        the ground state 
        up to 
        $\mathrm{2s.2p^{3}\,^{5}S^{o}}$~($4.18\,\mathrm{eV}$),
        while the high-excitation levels indicated in the plot span
        $\mathrm{2s^{2}.2p.3p\,^{1}P}$~($8.54\,\mathrm{eV}$)~up 
        to the ionisation limit~($11.26\,\mathrm{eV}$).
        The intermediate-excitation levels are those levels between them:
        $\mathrm{2s^{2}.2p.3s\,^{3}P^{o}}$~($7.49\,\mathrm{eV}$), 
        $\mathrm{2s^{2}.2p.3s\,^{1}P^{o}}$~($7.68\,\mathrm{eV}$),  
        and $\mathrm{2s.2p^{3}\,^{3}D^{o}}$~($7.95\,\mathrm{eV}$).
        Departure coefficients are shown
        for the standard model atom (\fig{fig:grotrian_red}).
        The impact of collapsing fine structure,
        and of removing neutral hydrogen impact excitation,
        are also shown.}
        \label{fig:departure}
    \end{center}
\end{figure}

In \fig{fig:departure} we illustrate 
departure coefficients 
across the \mtd~model atmosphere,
as predicted by the standard (reduced) model atom
used throughout this study.
There are broadly three groups of departure coefficients,
corresponding to levels of
low, intermediate, and high excitation potential;
\ion{C}{II}~belonging to the last group.
At solar metallicity,
the departures from LTE are driven by photon losses in 
the many strong \ion{C}{I} lines \citep[e.g.][]{1990A&A...237..125S}.
There is a population cascade, that starts in the 
levels of high excitation potential
and propagates down to the levels of intermediate excitation potential 
($\mathrm{2s^{2}.2p.3s\,^{3}P^{o}}$, $\mathrm{2s^{2}.2p.3s\,^{1}P^{o}}$, 
and $\mathrm{2s.2p^{3}\,^{3}D^{o}}$),
with levels of higher excitation potential underpopulating more.
The low-excitation levels are all highly populated,
and thus their populations are only marginally 
perturbed by this cascade (i.e.~their departure coefficients
stay close to unity).

We illustrate in \fig{fig:departure}
what happens when fine structure is collapsed in the model atom
(``No FS'').  
As discussed in \sect{methodatomreduced},
fine structure has a small impact on the statistical equilibrium.
It can 
be seen that in the 
regions $-2\lesssim\lgr\lesssim0$, the departure coefficients
are closer to unity in this case than in the standard case.
This is because collapsing fine structure leads to stronger lines,
that consequently form higher up in the solar atmosphere;
therefore, the departures from LTE start to occur
higher up in the solar atmosphere.
However, it can also be seen that even higher up,
$\lgr\lesssim-2$, the departure coefficients
are further from unity in this case than in the standard case.
This is because, when fine structure is collapsed,
the stronger lines have larger photon losses
and thus drive larger departure from LTE.

We also illustrate in \fig{fig:departure}
what happens when neutral hydrogen impact excitation
is neglected in the model atom
(``No FS / C+H exc.'').
Of all the collisional processes included in the model atom,
these have the largest impact on the statistical equilibrium,
generally acting to balance photon losses in the \ion{C}{I} lines.
When neutral hydrogen impact excitation is neglected, 
the departures from LTE set in much deeper in the atmosphere,
with significant departures already at
$\lgr\approx-0.5$.
It can also be seen that neutral hydrogen impact
excitation is responsible for strong
coupling within the different groups of levels
(of low, intermediate, and high excitation potential).

\subsection{3D/non-LTE abundance differences}
\label{resultseffects}

\begin{figure*}
    \begin{center}
        \includegraphics[scale=0.31]{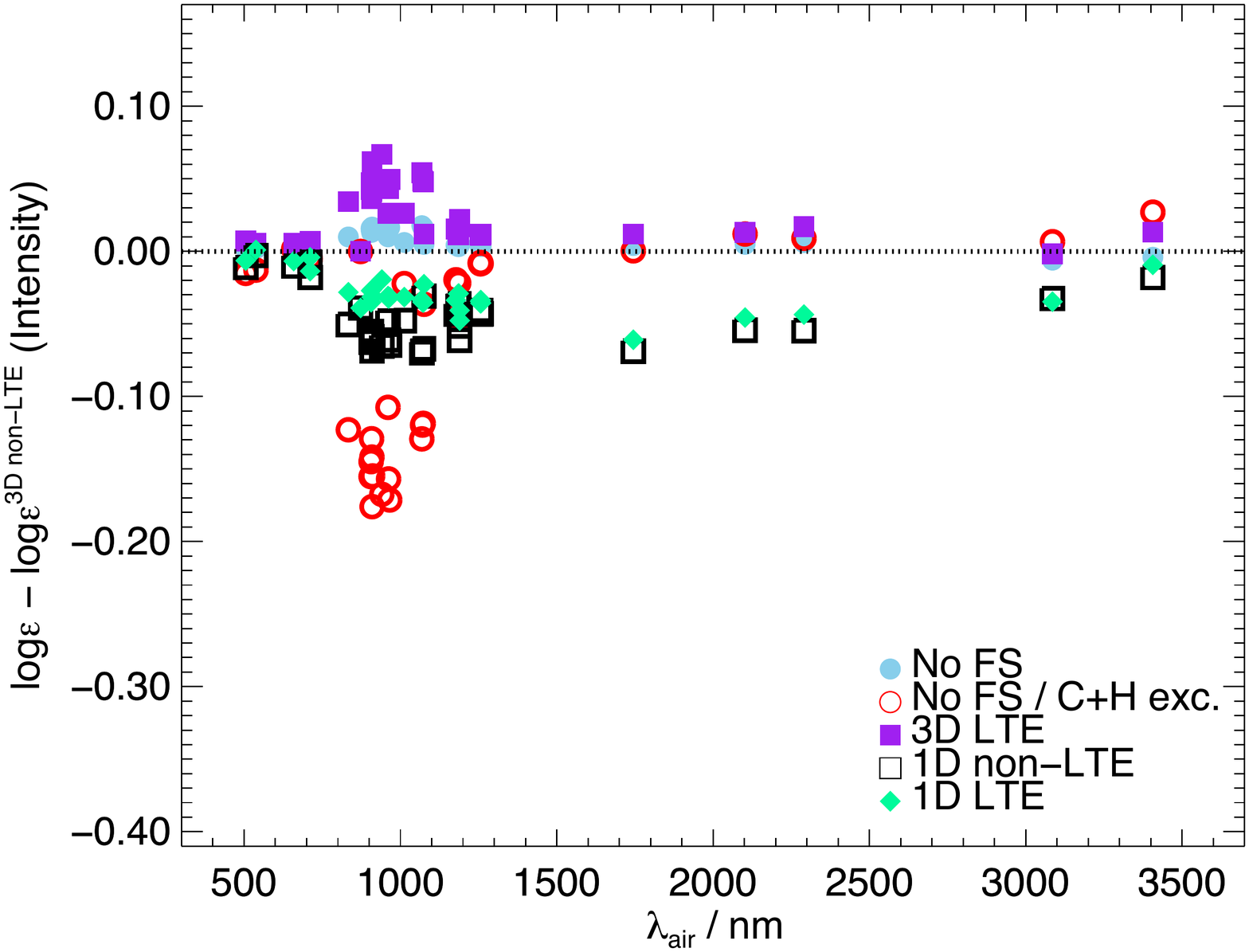}\includegraphics[scale=0.31]{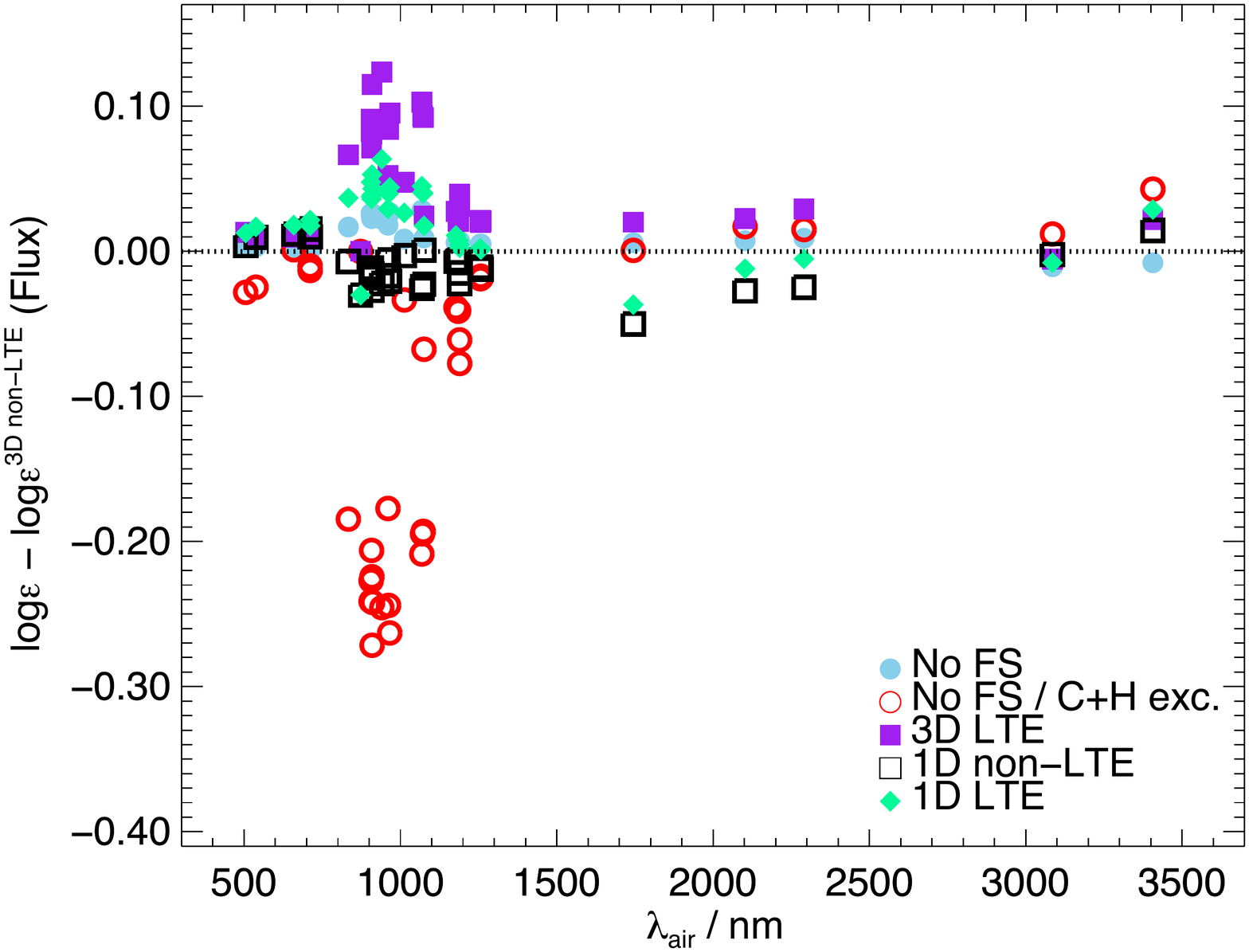}
        \includegraphics[scale=0.31]{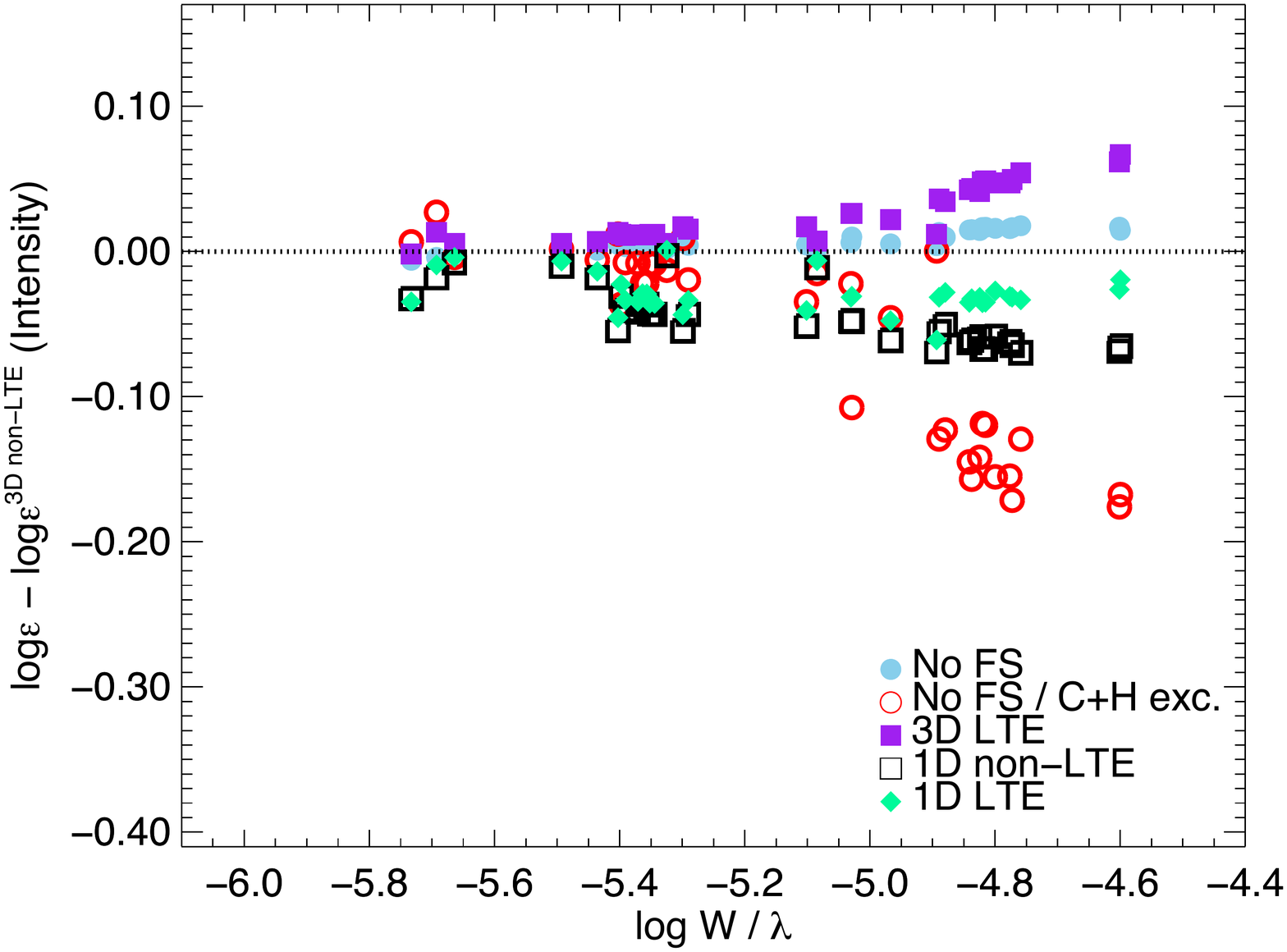}\includegraphics[scale=0.31]{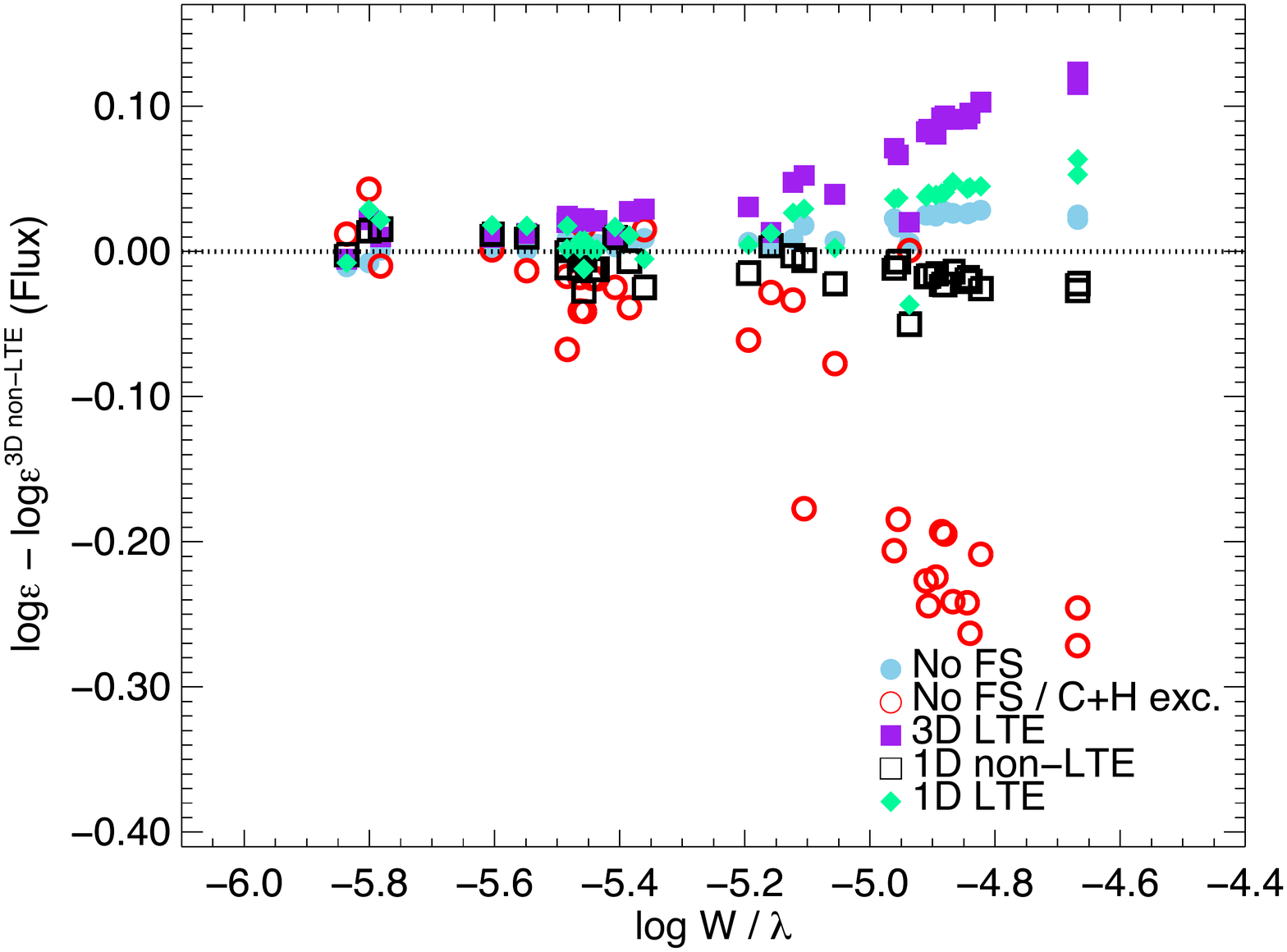}
        \includegraphics[scale=0.31]{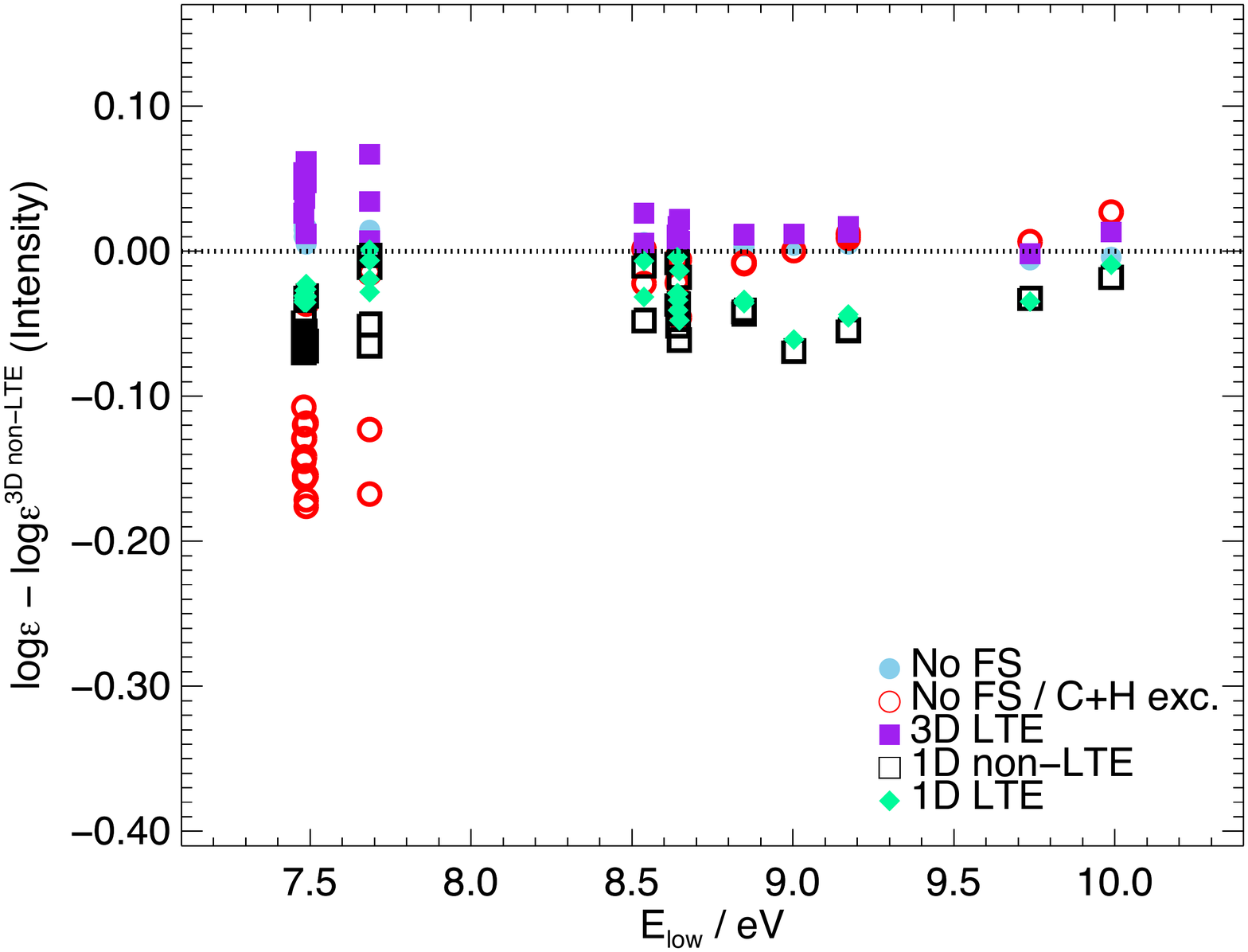}\includegraphics[scale=0.31]{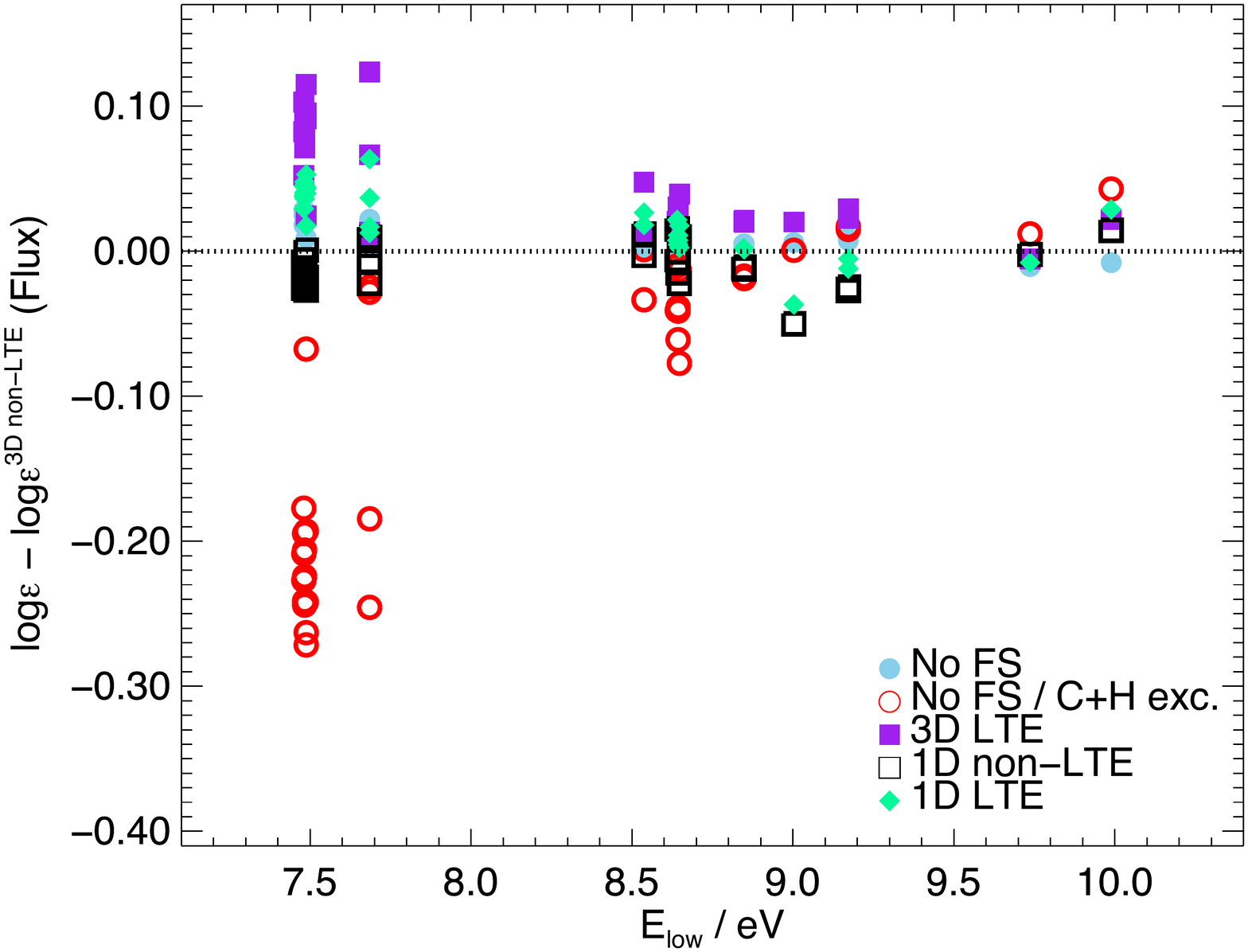}
        \caption{Abundance errors 
        (i.e.~abundance differences with respect to
        3D non-LTE)
        for the lines listed in \tab{tab:linelist}, when observed
        as disk-centre intensity (left column) 
        or as disk-integrated flux (right column).
        The [\ion{C}{I}] $872.7\,\nm$~line is not shown in these plots.
        Positive abundance errors imply negative abundance corrections.
        A fixed 3D non-LTE solar carbon abundance 
        of $8.43\,\mathrm{dex}$~was assumed.}
        \label{fig:aberr}
    \end{center}
\end{figure*}

To quantify the 3D non-LTE effects in the solar photosphere,
equivalent widths were calculated for all of
the lines listed in \tab{tab:linelist},
using different modelling approaches (e.g.~3D non-LTE, 1D LTE).
From these equivalent widths,
abundance differences with respect to 3D non-LTE were calculated,
by finding the abundance in, for example,
the 1D LTE approach, that is needed to match the
1D LTE equivalent width to the 
3D non-LTE equivalent width, for a fixed 3D non-LTE carbon abundance.
We hereafter refer to these abundance
differences as ``abundance errors'', in contrast
to the commonly used term ``abundance corrections''~that is understood
to represent abundance differences with respect to 1D LTE.
In \fig{fig:aberr} we illustrate these abundance errors
that are based on equivalent widths, in the solar atmosphere,
for the disk-centre intensity as well as 
for the disk-integrated flux.

In accordance with previous 1D non-LTE studies
\citep{2006A&amp;A...458..899F,2015MNRAS.453.1619A},
the 3D LTE versus 3D non-LTE abundance errors,
that quantify the non-LTE effects,
are generally positive in both the disk-centre intensity
and disk-integrated flux,
reaching up to around $0.1\,\dex$~in the latter.
In other words, the non-LTE abundance corrections 
tend to be negative. 
There are trends in the middle and lower panels
that indicate the departures from LTE are more severe 
for stronger \ion{C}{I} lines (larger $\log W / \lambda$),
and for \ion{C}{I} lines of intermediate excitation potential
($E_{\text{low}}\approx7.5$--$8.0\,\mathrm{eV}$).

The reason why the abundance corrections are usually
negative can be understood by looking
again at the departure coefficients in \fig{fig:departure}.
Levels of intermediate excitation potential do not depart
severely from their LTE populations,
whereas levels of high excitation potential
are severely underpopulated with respect to LTE.
Consequently, \ion{C}{I} lines of intermediate excitation potential
are subject to a source function effect
\citep[the ratio of the line source function to the Planck
function follows 
{$\beta_{\text{upper}}/\beta_{\text{lower}}$};][]{2003rtsa.book.....R}.
The line source function drops below the LTE expectation, 
and thus strengthens the \ion{C}{I} lines with respect to 
LTE.
Stronger lines are more susceptible to this effect, 
as their formation extends higher up into the atmosphere
where the departures from LTE are greater.

Similarly, \ion{C}{I} lines of high excitation potential
are also susceptible to the source function effect.
This is because higher levels show a larger
underpopulation with respect to LTE
(so {$\beta_{\text{upper}}/\beta_{\text{lower}}$}~still deviates
from unity).  For such lines, however,
there is also a competing opacity effect 
($\beta_{\text{lower}}$~drops below unity)
that acts to weaken the \ion{C}{I} lines with respect to LTE
(the line strength follows the line opacity).

The [\ion{C}{I}]~$872.7\,\nm$~line
does not depart from LTE. This is expected
because the lower and upper levels of the transition
are both of low-excitation and are thus both highly populated 
(since \ion{C}{I}~is the majority species in the solar photosphere).
They are therefore relatively insensitive to changes
in the populations in the intermediate- and
high-excitation \ion{C}{I}~levels.

We illustrate in \fig{fig:aberr} the sensitivity of the 
results to the treatment of fine structure,
by presenting errors associated with collapsing all fine structure
in the model atom (``No FS''). 
As anticipated (\sect{methodatomsimple}),
the abundance errors are close to zero.
They grow for stronger lines, which have significant
line formation higher up in the photosphere
where fine structure has a larger impact on the
statistical equilibrium
(as seen in \fig{fig:departure}).

We also illustrate in \fig{fig:aberr} the sensitivity of the 
results to neutral hydrogen impact excitation,
by presenting errors associated with neglecting these processes
(``No FS / C+H exc.'').
These processes have only
a small influence on the line strengths
of the optical lines considered here:
less than around $0.05\,\dex$~on the fluxes
in terms of abundances, at least in the Sun.
Their influence is much larger on the near infrared lines
(between $800\,\nm$~and $1100\,\nm$)
of intermediate excitation potential,
where including them tends to weaken the lines
considerably.  For these lines the effects can reach up to around
$0.3\,\dex$~on the fluxes in terms of abundances, at least in the Sun.
Neutral hydrogen impact excitation leads to closer
coupling between the lower and upper levels of the
near infrared lines (\sect{resultsdeparture}), and acts to
reduce the source function effect 
described above.

\fig{fig:aberr}~illustrates that 
the 1D non-LTE versus 3D non-LTE abundance errors,
that quantify the 3D effects,
are typically slightly smaller in magnitude than
the non-LTE effects.
However, these 3D effects go in the opposite direction
to the non-LTE effects: they are mainly negative
(corresponding to positive 3D abundance corrections,
such that the
1D non-LTE \ion{C}{I}~lines are too strong, relative to 3D non-LTE).
For the lines emergent at disk-centre,
the differences are typically around $0.05\,\dex$,
but can reach almost $0.1\,\dex$;
for the disk-integrated flux, the differences
are less severe (less than $0.05\,\dex$).
\fig{fig:aberr} shows that,
at disk-centre, the 3D effects are correlated with the 
line strengths ($\log W / \lambda$)
of the \ion{C}{I}~lines.
Reducing the microturbulence 
only slightly alleviates the 3D effects on the lines.
Rather, the 1D non-LTE versus 3D non-LTE abundance differences
are caused by differences
in the atmospheric stratification in 1D and in 3D;
in support of this,
we found that \mtd~non-LTE versus 3D non-LTE abundance corrections
are much less severe (less than $0.03\,\dex$~at disk-centre).

Since the vast majority of spectroscopic stellar analyses
today are based on 1D LTE, we also illustrate 
1D LTE versus 3D non-LTE abundance errors in
\fig{fig:aberr}.  
As discussed above, the 3D effects and 
the non-LTE effects tend to go in opposite directions,
at least for the \ion{C}{I} lines considered 
in this work.
Consequently, the 1D LTE versus 3D non-LTE abundance
errors are intermediate
between the negative 1D non-LTE ones (3D effects)
and the positive 3D LTE ones (non-LTE effects).
For the disk-centre intensity, 
they are always negative, which indicates 
that 3D effects are more important than non-LTE effects.
On the other hand, for the disk-integrated flux,
they tend to be positive, which indicates 
that non-LTE effects are more important than 3D effects.
The middle row, second column of \fig{fig:aberr}~reveals that
the differences are typically of the order $0.05\,\dex$~in the 
disk-integrated flux.

\subsection{Abundance trends}
\label{resultstrends}

\begin{table*}
\begin{center}
\caption{Equivalent widths in the disk-centre solar intensity and inferred solar carbon abundances from the different models: ``No FS'' refers to 3D non-LTE calculations using the model atom wherein all fine structure is collapsed; ``No FS / C+H exc.'' refers to 3D non-LTE calculations using the same model atom, but also excluding neutral hydrogen impact excitation of \ion{C}{I}. The uncertainties in the last column fold in uncertainties in the equivalent widths and oscillator strengths. The final row gives the weighted means, and standard error in the weighted mean, as discussed in \sect{resultstrends}. }
\label{tab:dcabund}
\begin{tabular}{c c c c c c c c c}
\hline
\multirow{2}{*}{$\lambda_{\text{air}}$} &
\multirow{2}{*}{$W / \mathrm{pm}$} &
\multicolumn{6}{c}{$\lgeps{C}$} &
\multirow{2}{*}{$\sigma_{\lgeps{C}}$} \\
 &
 &
3D non-LTE &
No FS &
No FS / C+H exc. &
3D LTE &
1D non-LTE &
1D LTE &
 \\
\hline
\hline
$  872.7121$ &
$      0.47$ &
$      8.45$ &
$      8.45$ &
$      8.45$ &
$      8.45$ &
$      8.41$ &
$      8.41$ &
$      0.05$ \\
\hline
$  505.2164$ &
$      4.05$ &
$      8.41$ &
$      8.41$ &
$      8.39$ &
$      8.42$ &
$      8.40$ &
$      8.40$ &
$      0.06$ \\
$  538.0335$ &
$      2.55$ &
$      8.43$ &
$      8.43$ &
$      8.41$ &
$      8.43$ &
$      8.43$ &
$      8.43$ &
$      0.06$ \\
$  658.7606$ &
$      1.79$ &
$      8.33$ &
$      8.33$ &
$      8.33$ &
$      8.33$ &
$      8.33$ &
$      8.33$ &
$      0.05$ \\
$  711.1467$ &
$      1.24$ &
$      8.31$ &
$      8.31$ &
$      8.30$ &
$      8.31$ &
$      8.31$ &
$      8.31$ &
$      0.05$ \\
$  711.3177$ &
$      2.55$ &
$      8.41$ &
$      8.41$ &
$      8.41$ &
$      8.42$ &
$      8.40$ &
$      8.40$ &
$      0.06$ \\
$  1075.397$ &
$      4.69$ &
$      8.49$ &
$      8.49$ &
$      8.45$ &
$      8.50$ &
$      8.46$ &
$      8.47$ &
$      0.06$ \\
$  1177.753$ &
$      6.30$ &
$      8.46$ &
$      8.47$ &
$      8.44$ &
$      8.48$ &
$      8.42$ &
$      8.43$ &
$      0.06$ \\
$  1254.949$ &
$      5.98$ &
$      8.51$ &
$      8.51$ &
$      8.50$ &
$      8.52$ &
$      8.47$ &
$      8.47$ &
$      0.06$ \\
$  1256.212$ &
$      6.35$ &
$      8.51$ &
$      8.51$ &
$      8.50$ &
$      8.52$ &
$      8.46$ &
$      8.47$ &
$      0.06$ \\
$  1256.904$ &
$      5.37$ &
$      8.46$ &
$      8.47$ &
$      8.46$ &
$      8.47$ &
$      8.42$ &
$      8.43$ &
$      0.06$ \\
$  1258.158$ &
$      5.83$ &
$      8.46$ &
$      8.46$ &
$      8.45$ &
$      8.47$ &
$      8.42$ &
$      8.42$ &
$      0.06$ \\
$  2102.316$ &
$      8.90$ &
$      8.47$ &
$      8.48$ &
$      8.48$ &
$      8.49$ &
$      8.42$ &
$      8.43$ &
$      0.06$ \\
$  3085.412$ &
$      5.56$ &
$      8.41$ &
$      8.41$ &
$      8.42$ &
$      8.41$ &
$      8.38$ &
$      8.38$ &
$      0.05$ \\
$  3406.491$ &
$      7.49$ &
$      8.47$ &
$      8.47$ &
$      8.50$ &
$      8.48$ &
$      8.46$ &
$      8.46$ &
$      0.05$ \\
\hline
\multicolumn{2}{c}{Weighted mean} &
$     8.437$ &
$     8.439$ &
$     8.433$ &
$     8.446$ &
$     8.411$ &
$     8.416$ &
$     0.016$ \\
\hline
\hline
\end{tabular}
\end{center}
\end{table*}

\begin{figure}
    \begin{center}
        \includegraphics[scale=0.31]{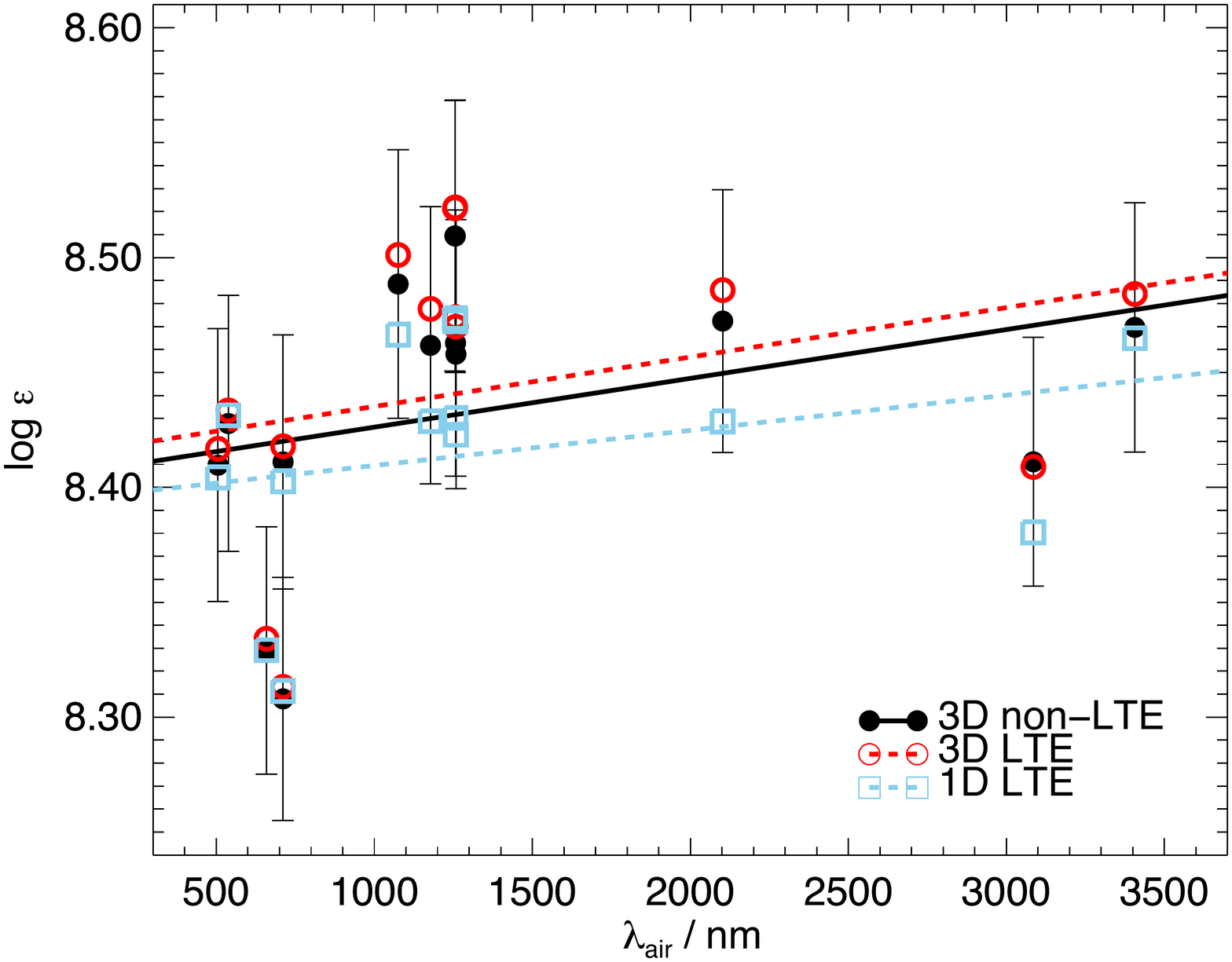}
        \includegraphics[scale=0.31]{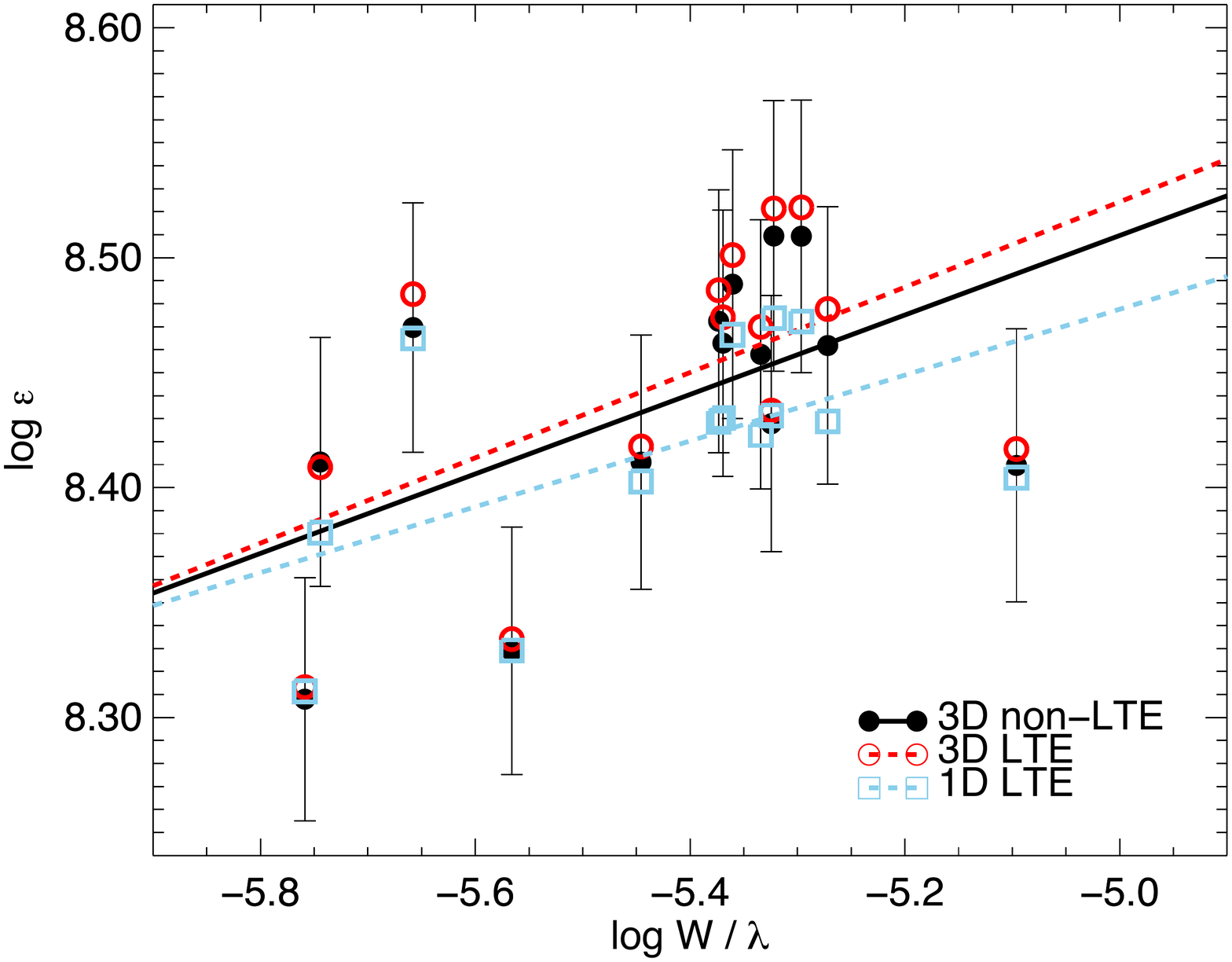}
        \includegraphics[scale=0.31]{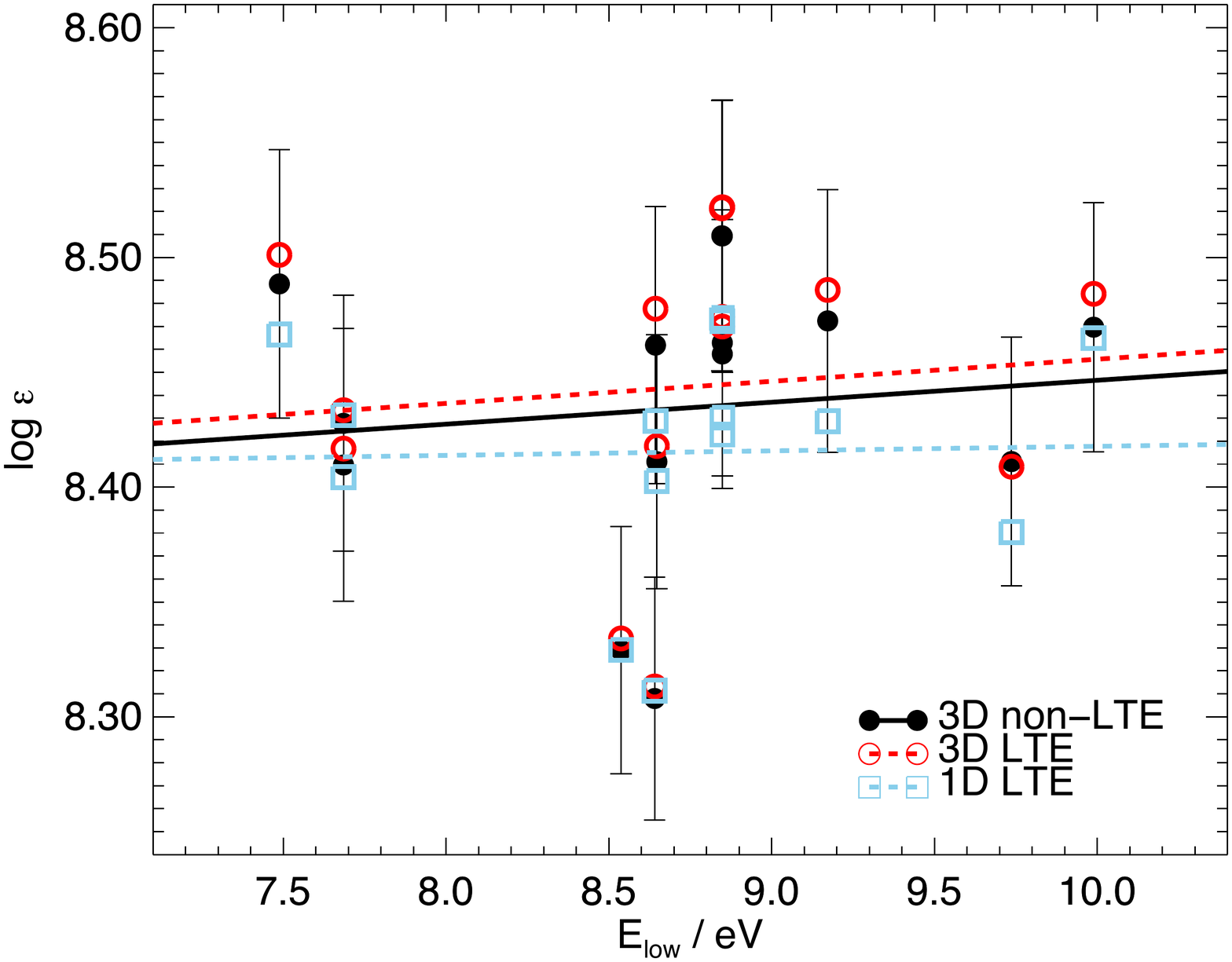}
        \caption{Inferred solar carbon abundances for different lines and
        models, based on disk-centre intensities.
        The [\ion{C}{I}] $872.7\,\nm$~line is not shown in these plots.
        Error bars are only shown for the 3D non-LTE model.
        Weighted linear fits are overdrawn.}
        \label{fig:trends}
    \end{center}
\end{figure}

In \tab{tab:dcabund}~we show our measured equivalent
widths for the [\ion{C}{I}]~line and 
\ion{C}{I}~lines measured in the disk-centre intensity spectrum,
together with the solar carbon abundances inferred
from different modelling approaches. 
Uncertainties in the equivalent widths
were assumed to be $5\%$, after consideration
of other published equivalent widths
\citep{1991A&amp;A...242..488G,
1993ApJ...412..431B,2005A&amp;A...431..693A,
2010A&amp;A...514A..92C}.
Uncertainties in the oscillator strengths (\tab{tab:linelist})
were also folded into the error budget,
simply assuming a one-to-one mapping of the error in $\log gf$~to
the error in $\lgeps{C}$, adopting the errors 
stipulated in NIST.

The mean abundances presented in the final row were 
then calculated after weighting individual lines $i$~in the usual way:
$w_{i}\propto1/\sigma_{i}^{2}$.
Under the assumption that all of the uncertainties are uncorrelated,
the standard error in the weighted means presented in
the final row is $0.02\,\dex$.
The forbidden [\ion{C}{I}]~$872.2\,\nm$~line was also included
in this weighted mean (i.e.~it was treated equally
to the permitted \ion{C}{I}~lines).
We discuss the mean abundance and its error further, in 
\sect{discussionabundance}.

The 3D non-LTE versus 3D LTE abundance 
differences in \tab{tab:dcabund}~are not very severe: typically
around $0.01\,\dex$. This can be contrasted with the 
values plotted in \fig{fig:aberr}, which can be in excess
of $0.10\,\dex$.  This is because the lines selected for determining
the solar carbon abundance are relatively insensitive
to departures from LTE.  Using the disk-centre intensity,
rather than the disk-integrated flux, further reduces the
impact of non-LTE effects.  This means that the inferred solar
carbon abundance should be less sensitive to any residual
modelling errors arising from the non-LTE model atom
and radiative transfer.

In \fig{fig:trends} we illustrate the inferred
solar carbon abundances as functions of different line parameters.
These plots are a useful check of possible systematic errors
in the models.  However, they have
less diagnostic power on different non-LTE models,
owing to the large scatter in the inferred line-by-line abundances.
For clarity, we only display the abundances
inferred from the 3D non-LTE, 3D LTE, and 1D LTE models
in these plots.

Using the 3D non-LTE model, consistent abundances are obtained
from the different \ion{C}{I}~lines, within a
standard deviation of $0.06\,\dex$.
This large dispersion may indicate
systematic errors in the models,
rather than uncertainties in the equivalent width 
measurements.
It is similar to 
the typical uncertainties in the individual oscillator strengths
(\tab{tab:linelist}), and could therefore be 
in part explained by errors in the oscillator strengths.

There is a slight trend in the inferred abundance
with wavelength, and a more
noticeable trend with line strength ($\log W / \lambda$)
that is significant at the $2\,\sigma$~level.
Larger abundances tend to be inferred from the stronger 
\ion{C}{I} lines than from the weaker ones.
The trend is mainly driven by the very low abundances
inferred from the two weak
\ion{C}{I} $658.8\,\nm$~and $711.1\,\nm$~lines,
as observed previously by \citet{2005A&amp;A...431..693A}.
To bring these two lines into agreement with the other lines,
their $\log gf$~values would need to be decreased by
about $0.1\,\dex$, or around $2\,\sigma$~if the errors in
\tab{tab:linelist}~are to be believed.
Improved \ion{C}{I}~oscillator strengths
would be highly desirable to help resolve this issue.

\subsection{Centre-to-limb variations}
\label{resultsclv}

\begin{figure*}
    \begin{center}
        \includegraphics[scale=0.31]{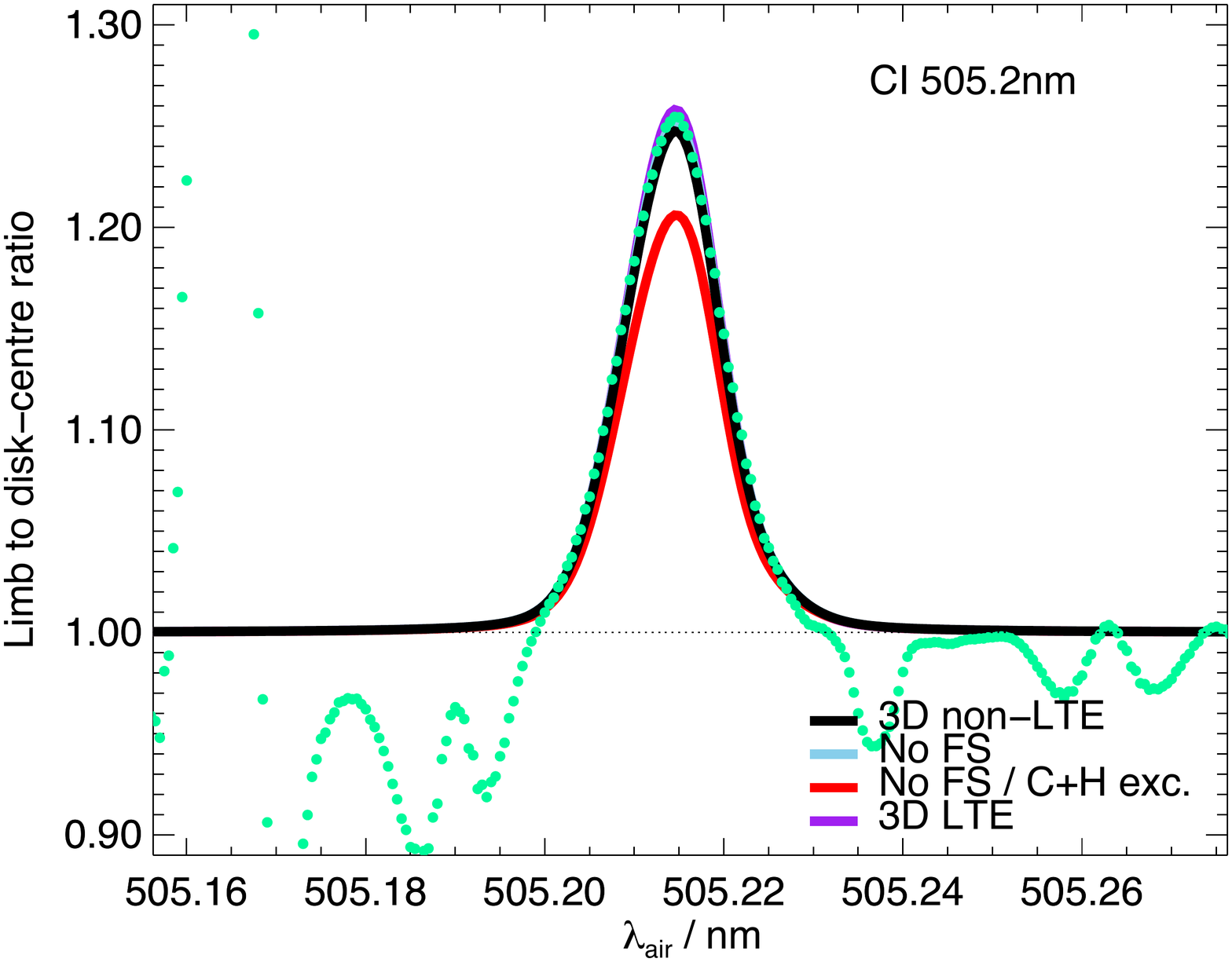}
        \includegraphics[scale=0.31]{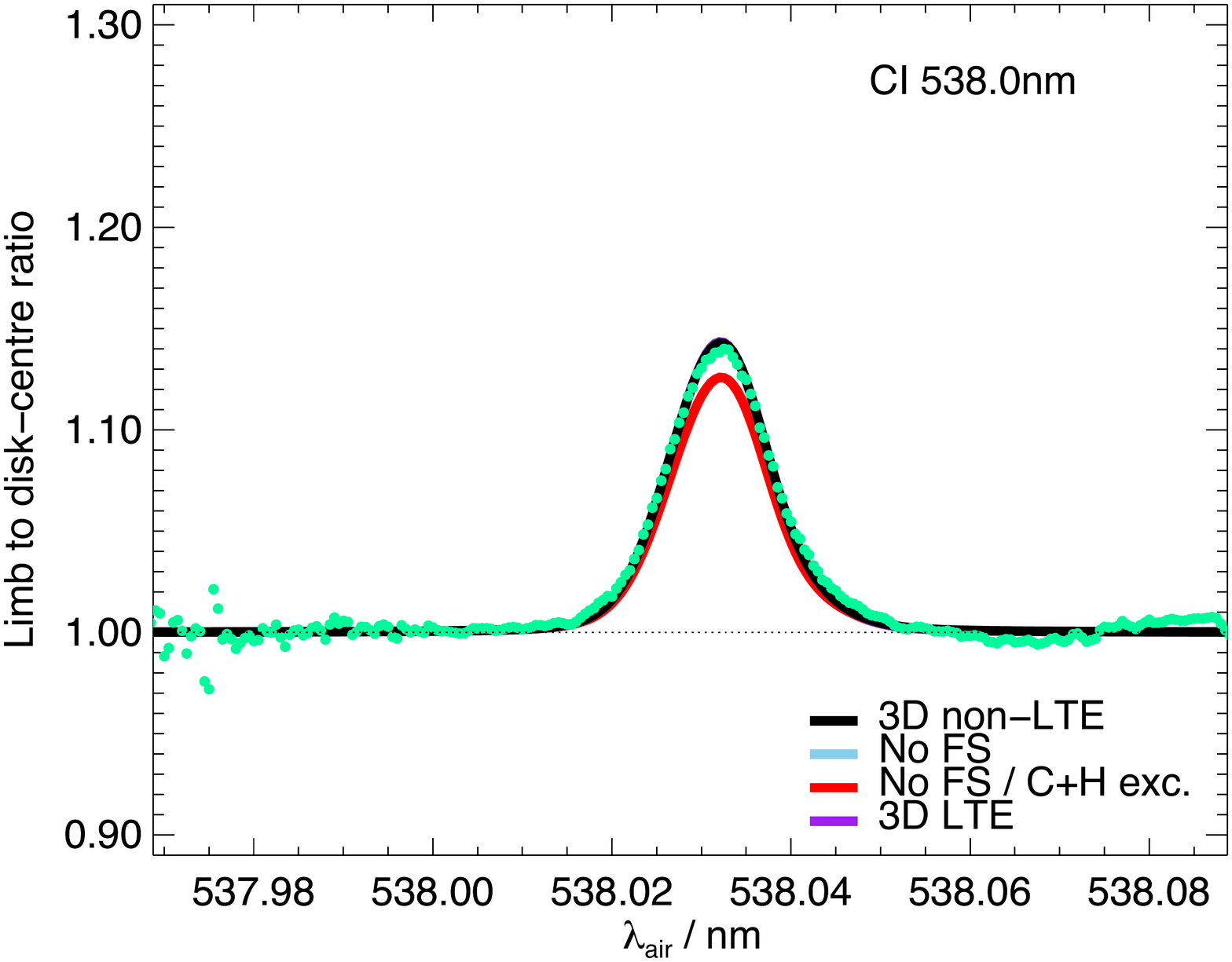}
        \includegraphics[scale=0.31]{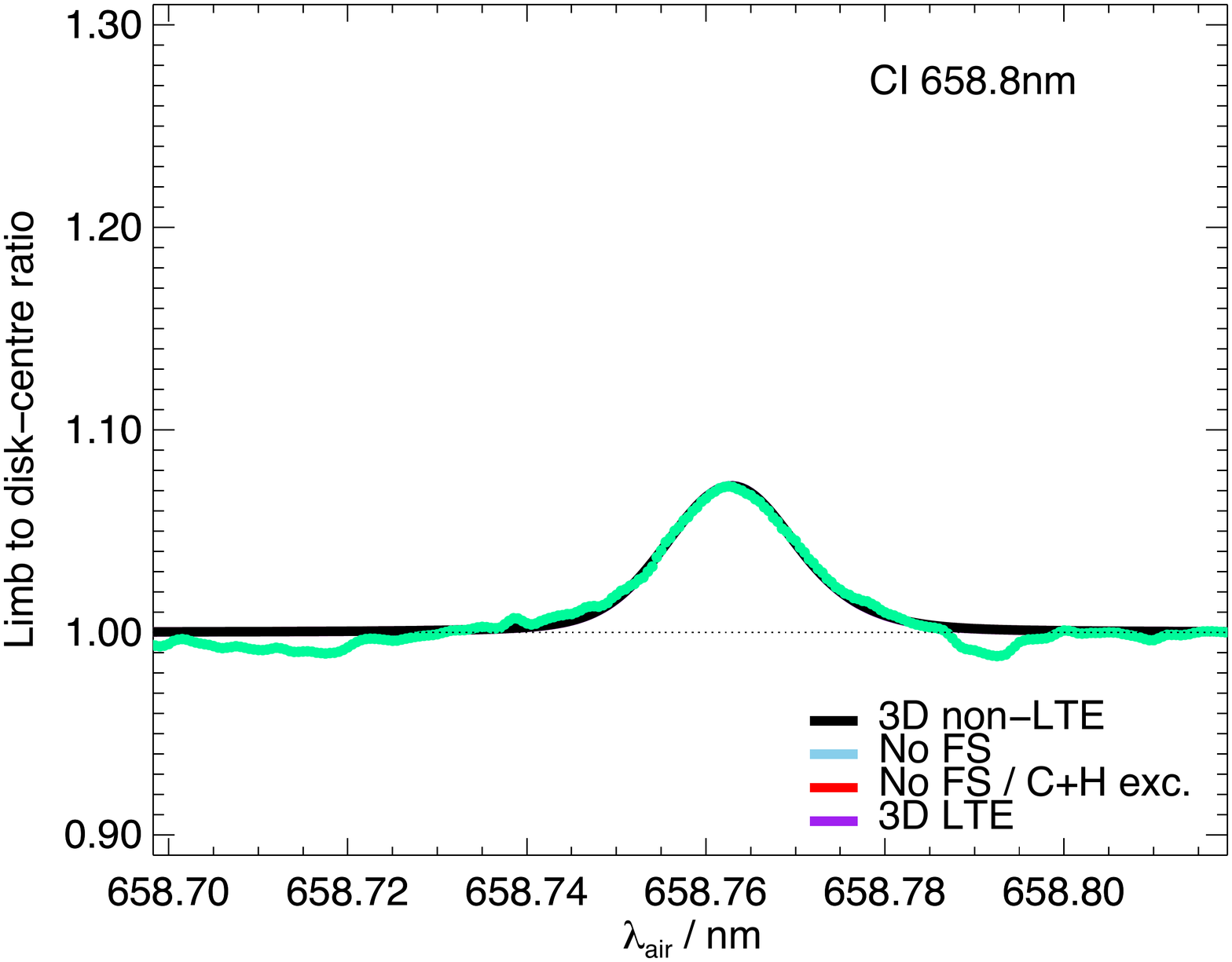}
        \includegraphics[scale=0.31]{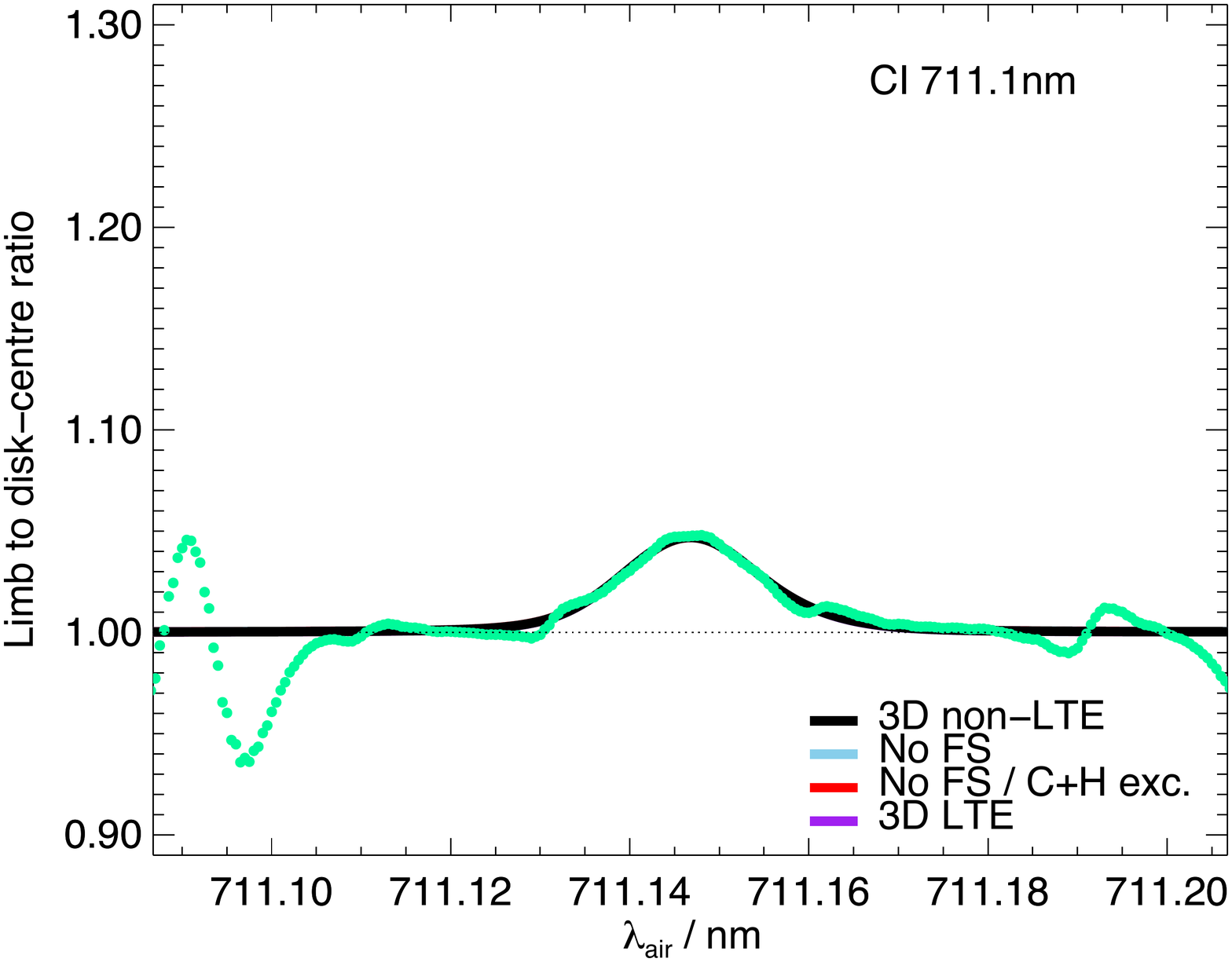}
        \caption{Comparison of the observed limb ($\mu\approx0.145$)
        to disk-centre ($\mu=1.0$) normalised intensity ratios
        with different model predictions, for different lines.
        The solar carbon abundances were set independently
        for each line and model using disk-centre intensity equivalent
        widths.  The solar carbon abundances and limb viewing angle
        were allowed to vary within their respective uncertainties,
        as described in \sect{resultsclv}.
        See also \fig{fig:clv2}.}
        \label{fig:clv1}
    \end{center}
\end{figure*}

\begin{figure*}
    \begin{center}
        \includegraphics[scale=0.31]{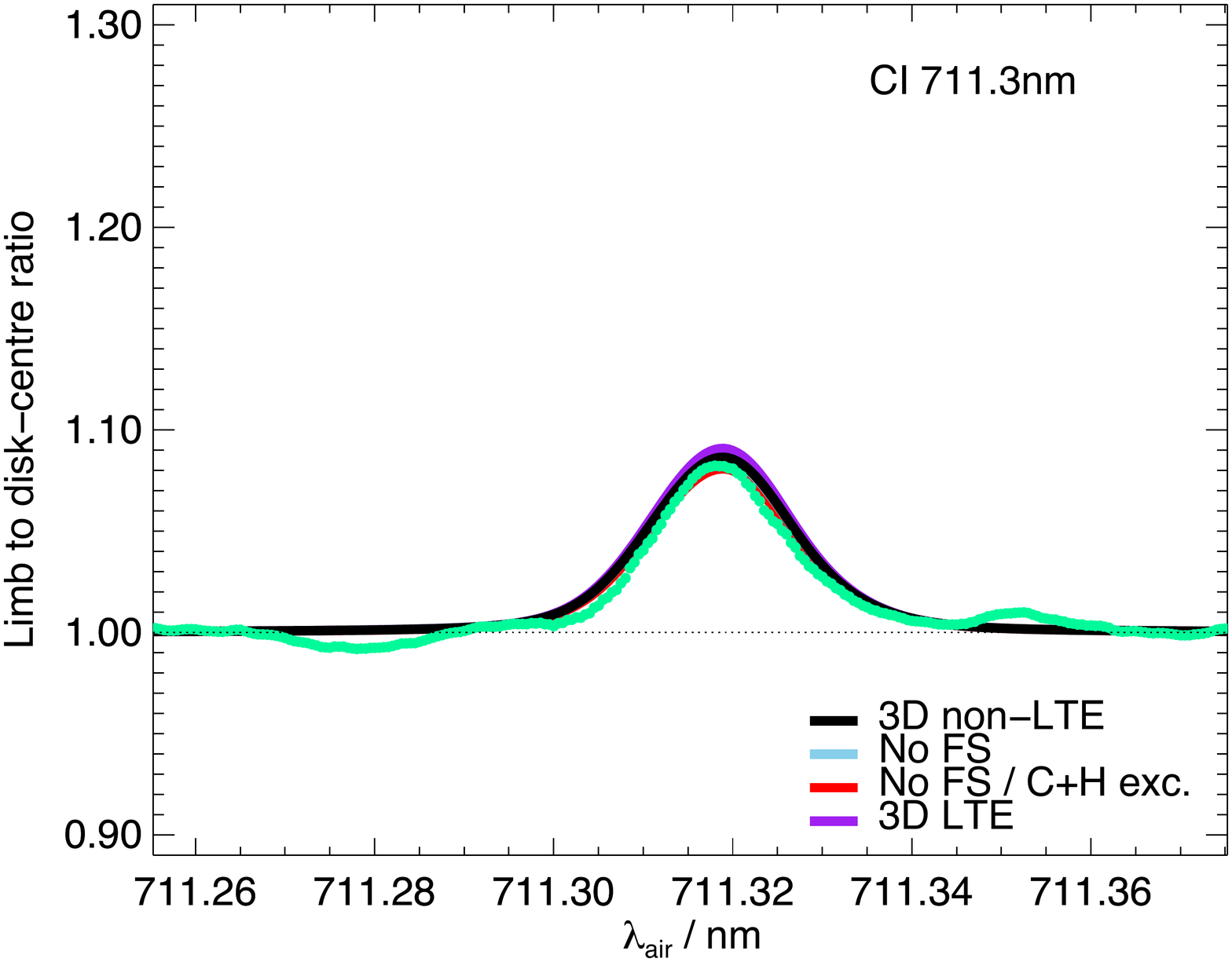}
        \includegraphics[scale=0.31]{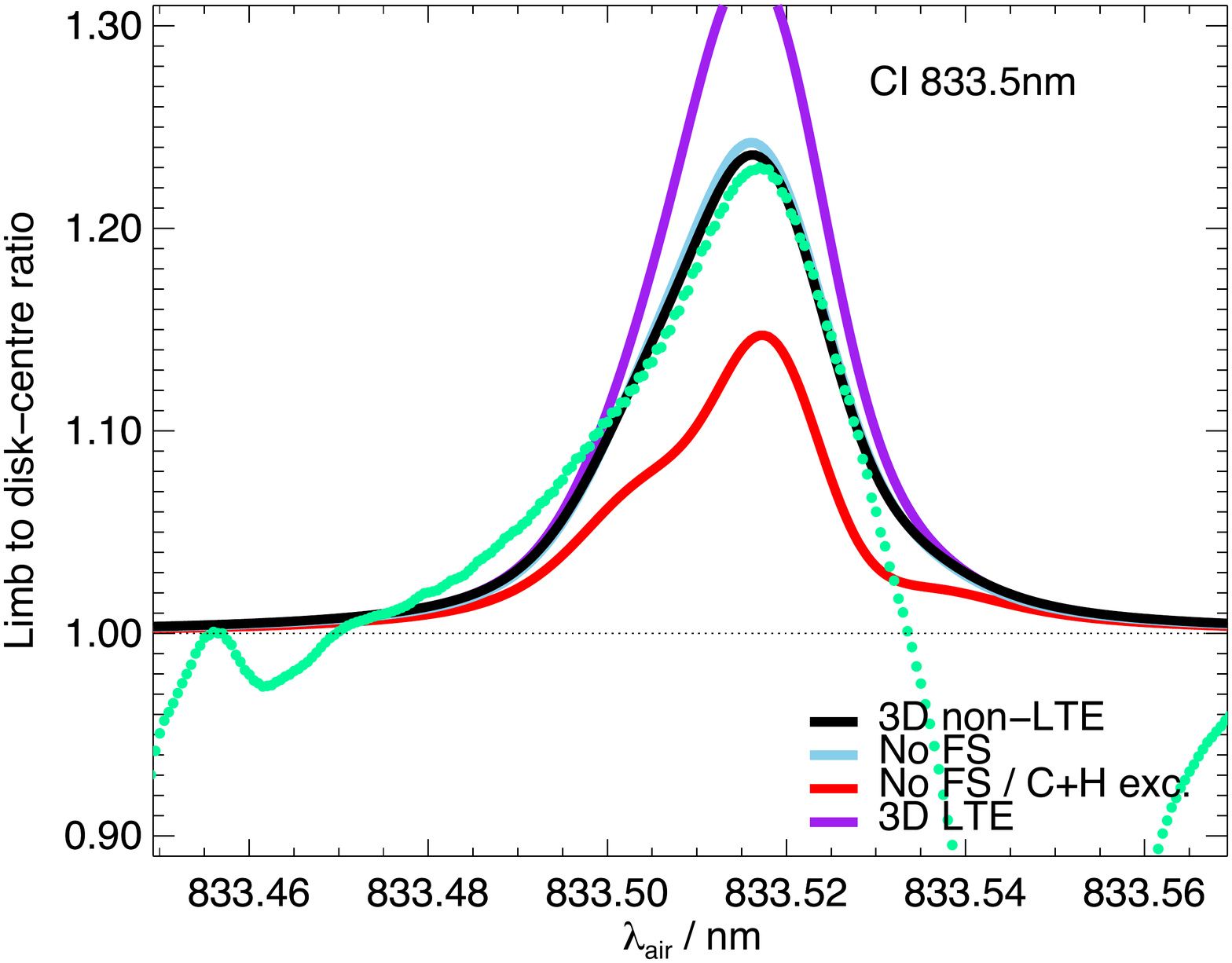}
        \includegraphics[scale=0.31]{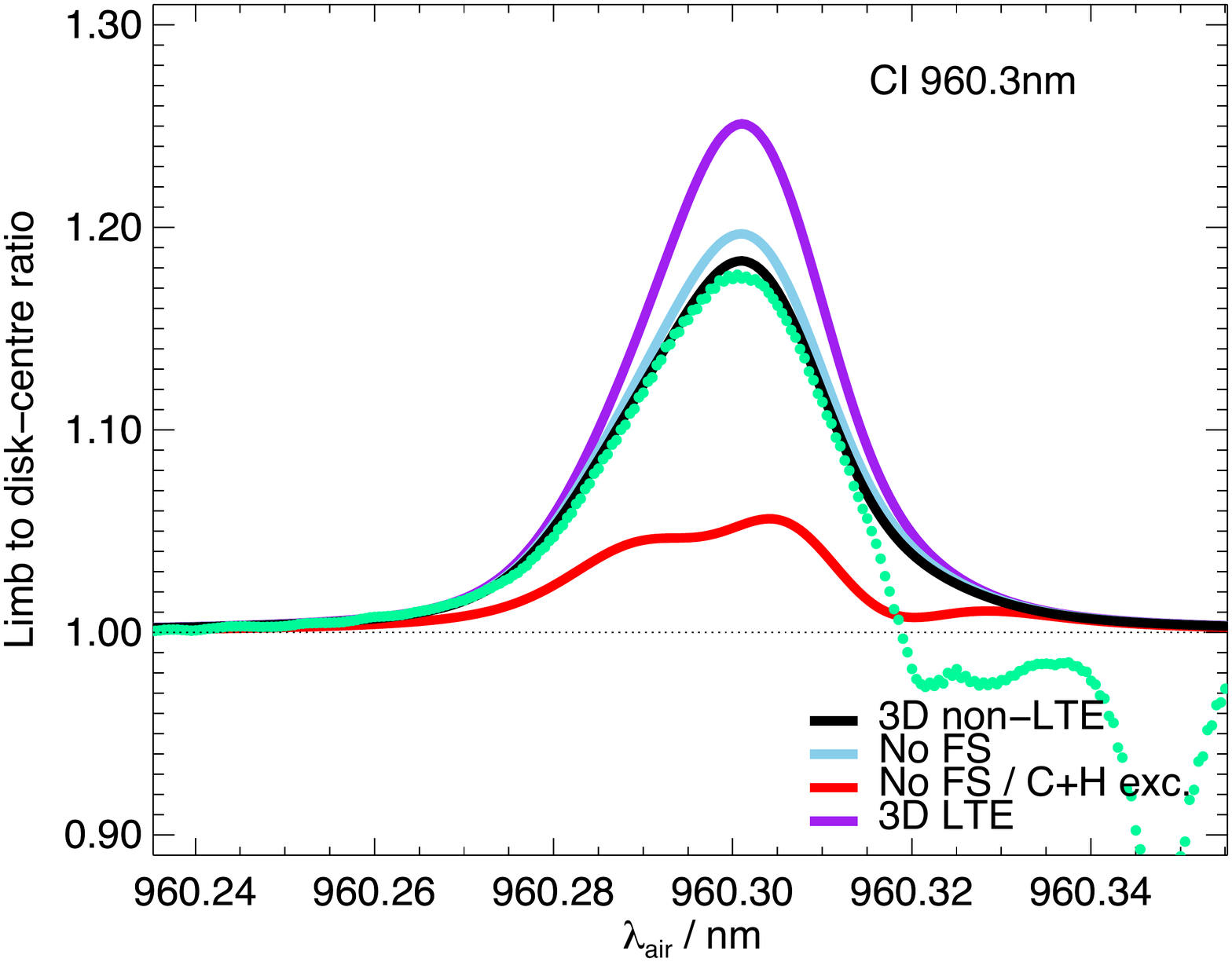}
        \includegraphics[scale=0.31]{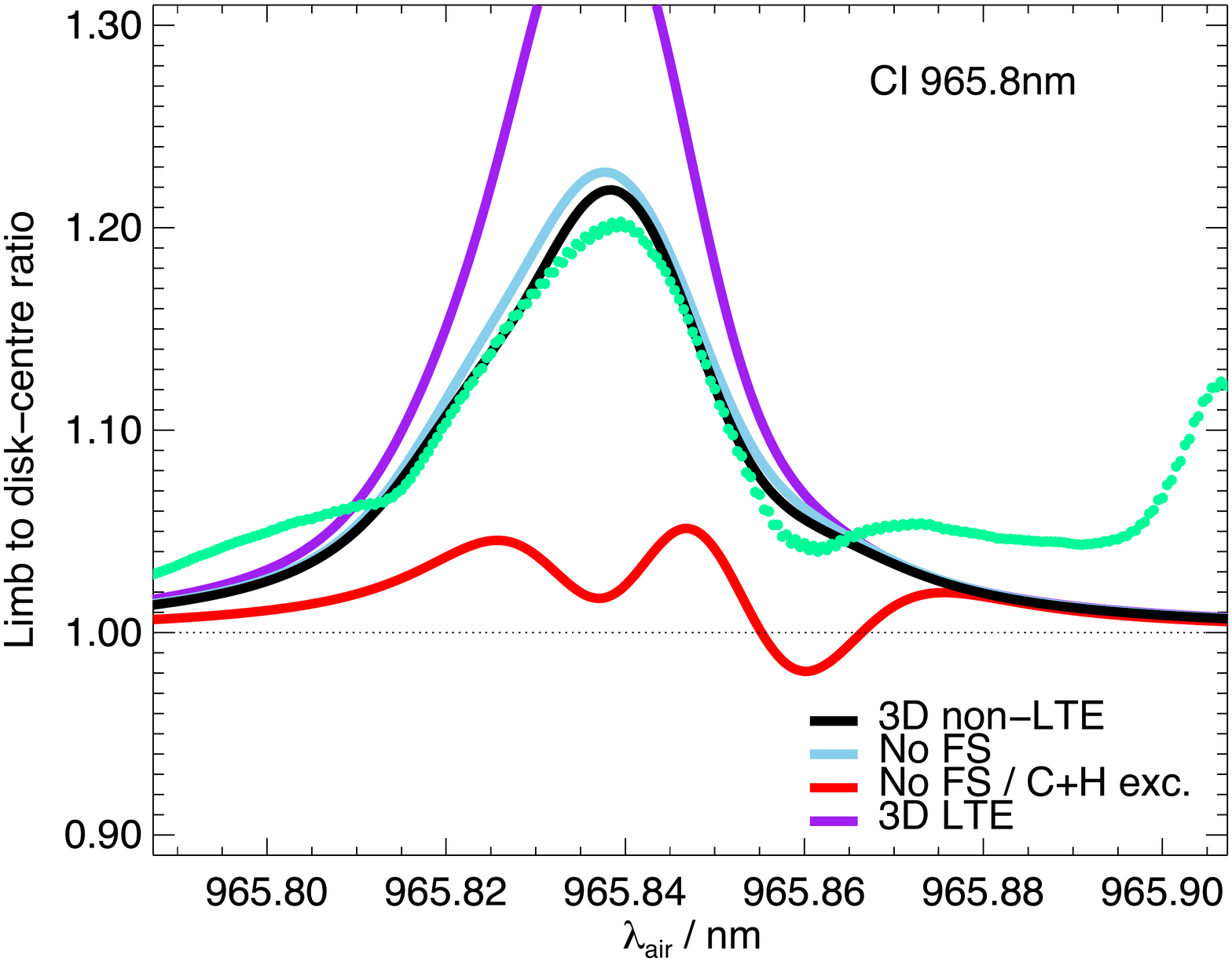}
        \caption{Comparison of the observed limb ($\mu\approx0.145$)
        to disk-centre ($\mu=1.0$) normalised intensity ratios
        with different model predictions, for different lines.
        The solar carbon abundances were set independently
        for each line and model using disk-centre intensity equivalent
        widths.  The solar carbon abundances and limb viewing angle
        were allowed to vary within their respective uncertainties,
        as described in \sect{resultsclv}.
        See also \fig{fig:clv1}.}
        \label{fig:clv2}
    \end{center}
\end{figure*}

The centre-to-limb variations of a
selection of \ion{C}{I}~lines (\sect{methodlinelist})
were studied using the high-resolution, high signal-to-noise
``SS3'' atlas of \citet{2015A&amp;A...573A..74S}
of the limb to disk-centre ratio atlas $R\left(\mu\right)$:
\phantomsection\begin{IEEEeqnarray}{rCl}
    r\left(\mu\right)&\equiv&
    I\left(\mu\right)/I^{\text{cont.}}\left(\mu\right)\, ,\\
    R\left(\mu\right)&\equiv&r\left(\mu\right)/r\left(1.0\right)\, .
\end{IEEEeqnarray}
\citet{2015A&amp;A...573A..74S} provide data
continuously from roughly $400\,\nm$~to $1000\,\nm$,
for $\mu=0.145$, and state that the uncertainty in the viewing angle
is at least $\sigma_{\mu}\approx0.01$.

The limb to disk-centre ratios predicted by~a)~the 3D non-LTE model, 
b)~the 3D non-LTE model but 
collapsing
all fine structure in the model atom 
(``No FS''),
c)~the 3D non-LTE model but neglecting 
collapsing
all fine structure in the model atom 
and neglecting 
neutral hydrogen impact excitation (``No FS / C+H exc.''),
and d) the 3D LTE model, were fit to the observations.
The 1D models were not considered here for two reasons.
First, we are mainly interested in comparing different 
non-LTE modelling approaches; this is best done using the 3D models.
Second, the 1D models 
are complicated by the requirements of extra broadening,
usually in the form of microturbulence and
macroturbulence fudge parameters.

Prior to performing the fits, the atlas was renormalised using 
clean regions close to each \ion{C}{I}~line.
When fitting each \ion{C}{I}~line,
the absolute wavelength calibration of the atlas was treated as
a free parameter. The viewing angle
was permitted to vary freely within the assumed
uncertainty of $\sigma_{\mu}=0.01$.
For each line and each model,
the solar carbon abundance was set to the values
inferred from analysing the disk-centre equivalent widths,
and was permitted to vary freely within an assumed
uncertainty of $0.05\,\dex$.
For most of the lines, the adopted disk-centre abundances
can be found in \tab{tab:dcabund}.
The \ion{C}{I} $833.5\,\nm$, $960.3\,\nm$, and $965.8\,\nm$~lines 
are missing from that table, as they are not
ideal solar carbon abundance indicators (\sect{methodlinelist}).
Therefore, for these lines the abundance 
analysis presented in \sect{resultstrends} was repeated, 
this time adopting disk-centre intensity equivalent widths
from \citet{2005A&amp;A...431..693A} for the 
\ion{C}{I} $960.3\,\nm$~line,
and from \citet{1993ApJ...412..431B} for the other two lines.

We illustrate the comparison of the model predictions to
the observations for the different 
\ion{C}{I}~lines in \fig{fig:clv1} and \fig{fig:clv2}.
Not all of the lines studied have diagnostic power:
the four different models predicted almost
identical limb to disk-centre ratios for the 
\ion{C}{I}~$658.8\,\nm$~and $711.1\,\nm$~lines
in the bottom row of \fig{fig:clv1},
and for these two lines
all four models are able to satisfactorily 
reproduce the observations.

As we mentioned in \sect{methodlinelist}, the
\ion{C}{I}~$711.3\,\nm$~is blended with a CN line,
that amount for around $7\%$~of the total
disk-centre intensity equivalent width. 
The centre-to-limb variation of the molecular blend
is different to that of the \ion{C}{I}~line, and
this is reflected in the results for the \ion{C}{I}~$711.3\,\nm$~in 
\fig{fig:clv2}: 
the limb to disk-centre ratios of the different models
are very similar, nevertheless, the ratio is 
slightly too strong compared to the observations
(the discrepancy is more noticeable, when this plot is
contrasted against the plot of 
the \ion{C}{I}~$711.1\,\nm$~line in \fig{fig:clv1}).
We stress again that our equivalent width and inferred abundances
in \tab{tab:dcabund} have taken this blend into account.

The results for the 
\ion{C}{I}~$833.5\,\nm$, $960.3\,\nm$,
and $965.8\,\nm$~lines 
in \fig{fig:clv2} clearly
favour the 3D non-LTE model over the 3D LTE model.
This is evidence for the need to take departures
from LTE into account, and for the accuracy of our model atom.
The quality of the fits is not perfect,
but this is understandable since 
the \ion{C}{I}~$960.3\,\nm$~line is blended with 
a CN~line in the far red wing,
while the \ion{C}{I}~$833.5\,\nm$~and 
\ion{C}{I}~$965.8\,\nm$~lines clearly suffer
from several small blends 
(including a \ion{Ti}{I}~line in the blue wing
of the \ion{C}{I}~$833.5\,\nm$~line),
which is why they are not considered reliable abundance diagnostics.

The \ion{C}{I}~$833.5\,\nm$, $960.3\,\nm$,
and $965.8\,\nm$~lines in \fig{fig:clv2} show that 
the standard model atom, in which fine structure is retained,
is able to fit the observations slightly better than the 
model atom in which fine structure is collapsed.
This is as expected from
the discussion in \sect{methodatomsimple},
with fine structure having a larger impact on the statistical
equilibrium higher up in the atmosphere.
Nevertheless, the difference between the two model
predictions for the limb to disk-centre ratios
is not very large.

Finally, it can be seen that 
neutral hydrogen impact excitation is indeed
important in the solar photosphere. 
This is clearly demonstrated by
the results for the \ion{C}{I}~$833.5\,\nm$, $960.3\,\nm$,
and $965.8\,\nm$~lines in \fig{fig:clv2},
and also by the
results for the \ion{C}{I}~$505.2\,\nm$~and
$538.0\,\nm$~lines in \fig{fig:clv1}.
This result is interesting, given that
the model adopts a new prescription for 
neutral hydrogen impact excitation
(``LCAO+Free''), as described in \sect{methodatomcollision}.

\section{Discussion}
\label{discussion}

\subsection{Comparison with previous studies}
\label{discussioncomparison}

In \sect{resultseffects}, we found 
that the non-LTE abundance corrections for \ion{C}{I} lines
tend to be negative in the solar photosphere.
This result can be compared with other 1D non-LTE analyses
\citep[e.g.][]{1990A&A...237..125S,
2005A&amp;A...431..693A,
2006A&amp;A...458..899F,
2010A&amp;A...514A..92C,
2015MNRAS.453.1619A}: these all predict
negative non-LTE abundance corrections, in agreement with
our result.

The most recent and comprehensive 1D non-LTE study 
is that of \citet{2015MNRAS.453.1619A}.
They constructed a large model \ion{C}{I} atom
that included fine structure. 
In their later study \citep{2016MNRAS.462.1123A}, they updated
their model atom to include the more reliable 
BSR electron impact excitation and ionisation data
\citep[][]{2006CoPhC.174..273Z}, that was also adopted here,
this had only a small impact on their results
(Alexeeva, priv.~comm.). The main difference 
between their model and that adopted here is the prescription for 
neutral hydrogen impact excitation:
they adopted the Drawin recipe
\citep{1968ZPhy..211..404D,1969ZPhy..225..483D,
1984A&amp;A...130..319S,1993PhST...47..186L},
whereas the model used here adopted the ``LCAO+Free'' approach
described in \sect{methodatomcollision}.

Comparing Fig.~2 of
\citet{2015MNRAS.453.1619A} with
\fig{fig:departure} here, the departure coefficients
between the two studies are in good qualitative agreement.
The low-excitation levels stay close to unity,
the intermediate-excitation levels are 
slightly overpopulated with respect to LTE,
and the high-exctation levels underpopulate with respect to LTE.
As illustrated in \fig{fig:departure} here,
the extent of the departures from LTE are very sensitive
to neutral hydrogen impact excitation,
and this can explain for example the larger overpopulation
of the intermediate-excitation levels compared to ours.

\subsection{Solar carbon abundance}
\label{discussionabundance}

In \sect{resultstrends}~the 
solar carbon abundance was inferred from different 
\ion{C}{I}~lines as a diagnostic for the 3D non-LTE 
line formation models. We can also use this analysis
to comment on the solar carbon abundance;
a fully consistent 3D non-LTE determination
has not been presented in the literature before.

Our derived abundance is $\lgeps{C}=8.44\pm0.02$.
The value is the weighted mean from the fully consistent
3D non-LTE analysis presented in \tab{tab:dcabund}.
The uncertainty was calculated in a way similar
to the way explained in 
Sect.~4.1 of \citet[][]{2015A&amp;A...573A..25S}:
the systematic error arising from the non-LTE modelling was
estimated by taking half the difference
between the 3D LTE and 3D non-LTE results 
($0.004\,\dex$),
and the systematic error arising from the 3D modelling was
estimated by taking half the difference
between the 1D non-LTE and 3D non-LTE results 
($0.013\,\dex$).
These were combined in quadrature ($0.014\,\dex$).
This systematic error was combined in quadrature with
the standard error in the weighted mean
($0.016\,\dex$), to get the final uncertainty 
($0.02\,\dex$).

Our derived value is 
consistent with
the current standard value 
of $\lgeps{C}=8.43\pm0.05$~\citep{2009ARA&amp;A..47..481A}.
In fact, it is in good agreement
with the abundances inferred from
each of the five different diagnostics employed in that work:
the [\ion{C}{I}]~$872.7\,\nm$~line ($8.41\,\dex$),
$16$~\ion{C}{I}~lines ($8.42\,\dex$),
CH\,$\Delta\varv=1$~lines ($8.44\,\dex$),
CH\,$\mathrm{A-X}$~lines ($8.43\,\dex$),
and C$_{2}$\,$\mathrm{Swan}$~lines ($8.46\,\dex$).

The [\ion{C}{I}]~line is insensitive to departures from LTE
in the solar photosphere, and is relatively 
insensitive to the model atmosphere 
\citep[see][Table 2]{2009ARA&amp;A..47..481A}.
Most of the differences between the result of
\citet{2009ARA&amp;A..47..481A},
of $8.41\,\dex$,
and our result of $8.45\,\dex$,
can be attributed to
their larger oscillator strength:
$\log gf=-8.136$, compared to our $\log gf=-8.165$,
which leads to a $0.03\,\dex$~difference in inferred abundance.

Our derived carbon abundance is
significantly lower than the value
of \citet{2010A&amp;A...514A..92C,2011SoPh..268..255C},
who inferred $\lgeps{C}=8.50\pm0.06$.
Their abundances were based on a 3D LTE analysis
of the [\ion{C}{I}]~$872.7\,\nm$~line 
and $44$~\ion{C}{I}~lines;
as with \citet{2009ARA&amp;A..47..481A}, 
they applied negative 1D non-LTE versus 1D LTE
abundance corrections for the \ion{C}{I} lines.

The main reason for the discrepancy
between our result 
and that of \citet{2010A&amp;A...514A..92C,2011SoPh..268..255C}
can be traced to differences in the equivalent widths.
Their equivalent widths tend to be systematically larger than those 
adopted here.  
Adopting their equivalent widths, and lines for 
that they assigned a rank of $1$, 
we obtain a solar carbon abundance that is around 
$0.07\,\dex$~larger than our derived value:
that is, $\lgeps{C}\approx8.51$,
in good agreement with their result.
Looking carefully at the line shapes
as well as databases of known lines in the solar spectrum,
one can see that around two thirds of the \ion{C}{I}~lines used by 
\citet{2010A&amp;A...514A..92C,2011SoPh..268..255C}
are blended, and are therefore 
not suitable abundance indicators, tending to bias
the inferred carbon abundance upwards. 
Including blended lines also results in a significantly
larger scatter in the results,
with a standard deviation in the line-by-line
carbon abundances of $0.11\,\dex$:
compare Fig.~3 of \citealt{2005A&amp;A...431..693A}
with Fig.~3 of \citealt{2010A&amp;A...514A..92C}, noting the difference
in scales.

\subsection{Solar C/O ratio}
\label{discussionratio}

We briefly comment on the solar C/O ratio,
having also recently presented a similar 3D non-LTE analysis
of the \ion{O}{I}~$777\,\nm$~triplet
\citep{2018A&A...616A..89A}.
In that work, we obtained
$\lgeps{O}=8.69\pm0.03$, in agreement with the 
current standard value~\citep{2009ARA&amp;A..47..481A}.
Combining this with our value of 
$\lgeps{C}=8.44\pm0.02$, and assuming that these two uncertainties
are uncorrelated, a solar C/O ratio of 
$0.56\pm0.05$~is inferred.
This is consistent with the 
C/O ratios of \citet{2009ARA&amp;A..47..481A},
and also of \citet{2011SoPh..268..255C},
namely $0.55\pm0.09$~and 
$0.55\pm0.12$~respectively.

There remains a discrepancy with
recent in-situ measurements of the fast solar wind
from polar coronal holes: $0.67\pm0.06$~\citep{2016ApJ...816...13V}.
We caution however that the C/O ratio in the solar wind
may not be reflective of that in the solar photosphere
owing to the First Ionisation Potential (FIP)
effect and related fractionation effects
\citep[e.g.][]{2016MNRAS.463....2S}.
Nevertheless, current models suggest that such effects have 
only a small impact on the C/O ratio
\citep[e.g.][Table 4]{2015LRSP...12....2L},
reflecting the similar ionisation potentials
of neutral carbon and neutral oxygen.

On the other hand, this C/O ratio
is larger than that measured in
solar neighbourhood B-type stars.
\citet{2012A&A...539A.143N} infer a larger oxygen
abundance and a smaller carbon abundance from B-type stars,
and infer a C/O ratio of $0.37\pm0.05$,
significantly lower than our solar estimate. 
The C/O ratio at the solar surface should roughly reflect that
of the present-day local cosmos, and consequently
that in short-lived B-type stars, because
to first order carbon and oxygen are expected to be similarly
affected by thermal diffusion, gravitational settling, 
and radiative acceleration, and should have had similar
Galactic chemical enrichment histories 
\citep[e.g.][and references therein]{2009ARA&amp;A..47..481A}.
As discussed in Sect.~7.5 of \citet[][]{2012A&A...539A.143N},
a higher C/O ratio in the Sun compared 
to in nearby B-type stars is probably a sign that the
Sun was born at a location closer to the centre of the Galaxy
and later migrated outwards to its present location.

\section{Conclusion}
\label{conclusion}

We have presented a new model atom for 
non-LTE analyses of \ion{C}{I}~line formation in late-type stars.
We used the model atom together with detailed
3D non-LTE radiative transfer calculations
across a 3D hydrodynamic model solar atmosphere,
and analysed the solar spectrum.
The model atom successfully reproduced the observed
limb to disk-centre ratios of various \ion{C}{I}~lines, and
conclusively ruled out 3D LTE as a viable model.
We also presented the first consistent 3D non-LTE solar carbon abundance
determination. Our derived value
is $\lgeps{C}=8.44\pm0.02$.
The new solar carbon abundance is consistent with the
current standard
value of $\lgeps{C}=8.43\pm0.05$~presented in
\citet{2009ARA&amp;A..47..481A}.
Combined with an earlier 3D non-LTE solar oxygen abundance
of $\lgeps{O}=8.69\pm0.03$, a solar
C/O ratio of $0.56\pm0.05$~was inferred,
again consistent with the current standard value of 
$0.55\pm0.09$.

The true nature of inelastic collisions with neutral
hydrogen, and their role on non-LTE stellar spectroscopy,
is a decades-old problem. The main improvement in our model atom
compared with previous studies is 
the more realistic, ``LCAO+Free'' description of 
neutral hydrogen impact excitation.
This approach was first motivated in our earlier study of 
the solar centre-to-limb variation 
of the high excitation potential \ion{O}{I}~$777\,\nm$~triplet
\citep{2018A&A...616A..89A}.
As in that work, the 3D non-LTE line formation model in this work
successfully reproduced the 
limb to disk-centre ratios of various 
\ion{C}{I} lines, only when neutral hydrogen impact excitation
was included. This suggests that the
``LCAO+Free'' recipe is a suitable approach
for non-LTE modelling of neutral lines in late-type stars,
although we cannot rule out some unknown, compensating systematic 
errors in our models.
Since neglecting neutral hydrogen impact excitation can
impart abundance errors
of up to $0.3\,\dex$~on disk-integrated fluxes,
at least for \ion{C}{I} lines in the Sun, it is important
to take these collisional processes into account.


\begin{acknowledgements}
We thank the referee, Lyudmila Mashonkina,
for carefully reading
and providing helpful suggestions on the manuscript,
and in particular for insight and 
comments which helped identify a numerical
issue with the code and thus greatly improved the paper. 
We also thank Sofya Alexeeva for answering in detail
our various questions about her analysis,
and Karin Lind and Amanda Karakas for comments on the manuscript.
AMA acknowledges funds from the Alexander von Humboldt Foundation in the
framework of the Sofja Kovalevskaja Award endowed by the Federal Ministry of
Education and Research.
PSB acknowledges support from the Swedish Research Council and the 
project grant ``The New Milky Way'' from the Knut and 
Alice Wallenberg Foundation.
Funding for the Stellar Astrophysics Centre is provided by The Danish
National Research Foundation (grant DNRF106).
MA gratefully acknowledges funding from
the Australian Research Council
(grants DP150100250 and FL110100012).
Some of the computations were performed on resources provided by 
the Swedish National Infrastructure for Computing (SNIC) at 
the Multidisciplinary Center for Advanced Computational Science (UPPMAX) 
and at the High Performance Computing Center North (HPC2N)
under project SNIC2018-3-465.
This work was supported by computational resources provided by 
the Australian Government through the 
National Computational Infrastructure (NCI)
under the National Computational Merit Allocation Scheme.

\end{acknowledgements}


\bibliographystyle{aa} 
\bibliography{/Users/ama51/Documents/work/papers/allpapers/bibl.bib}


\label{lastpage}
\end{document}